\documentclass[jmp, 12pt]{revtex4-1}

\usepackage{graphicx}
\usepackage[caption=false]{subfig}

\usepackage{braket}
\usepackage{color}
\usepackage{amsmath}
\usepackage{amssymb}
\usepackage{natbib}

\begin{document}

\title{Semiclassical asymptotics of the Aharonov-Bohm interference process}

\author{Stefan G. Fischer}
\author{Clemens Gneiting}
\author{Andreas Buchleitner}
\affiliation{Physikalisches Institut, Universit\"at Freiburg, Hermann-Herder-Stra\ss e 3, D-79104 Freiburg, Germany}

\begin{abstract}
In order to determine the origin of discontinuities which arise when the semiclassical propagator is employed to describe an infinitely long and infinitesimally thin solenoid carrying magnetic flux, we give a systematic derivation of the semiclassical limit of the motion of an otherwise free charged particle. Our limit establishes the connection of the quantum mechanical canonical angular momentum to its classical counterpart.
Moreover, we show how a picture of Aharonov-Bohm interference of two half-waves acquiring Dirac's magnetic phase when passing on either side of the solenoid emerges from the quantum propagator, and that the typical scale of the resulting interference pattern is fully determined by the ratio of the angular part of Hamilton's principal function to Planck's constant. The semiclassical propagator is recovered in the limit when this ratio diverges. We discuss the relation of our results to the whirling-wave representation of the exact propagator.
\end{abstract}

\pacs{03.65.Sq, 03.65.Vf}
\maketitle
\section{\label{sec:intro}Introduction}

In a series of thought experiments, Aharonov and Bohm demonstrated that charged particles are influenced by electromagnetic potentials, even if their wave function vanishes wherever the associated electromagnetic field is nonzero~\cite{AHARONOVBOHM}. A controversial debate on whether such effects existed has since been resolved by their experimental verification~\cite{PESHKINTONOMURA}. 
Phase shifts due to the presence of inaccessible magnetic flux have been confirmed in electron microscope experiments~\cite{CHAMBERS,MOELLENSTEDTI,MOELLENSTEDTII} as well as by means of electron holography~\cite{TONOMURA}, and with the same technique where the flux was additionally shielded by a superconducting layer, to further minimize the influence of stray magnetic fields~\cite{TONOMURAII}. 
The simulation of Aharonov-Bohm phases in cold atom systems trapped in optical lattices~\cite{JAKSCH} is now experimentally feasible~\cite{AIDELSBURGER}, and the effect plays an important role in mesoscopic electron-optical systems~\cite{JI,NEDERII,NEDERIII}, where
an interesting manifestation for charged identical particles~\cite{BUETTIKER,DERAEDT,BUETTIKERII,BUETTIKERIII} has recently been confirmed in an electronic 
Hanbury Brown and Twiss interferometer~\cite{NEDER}.

\begin{figure}
\centering
  \includegraphics[width=0.9\textwidth]{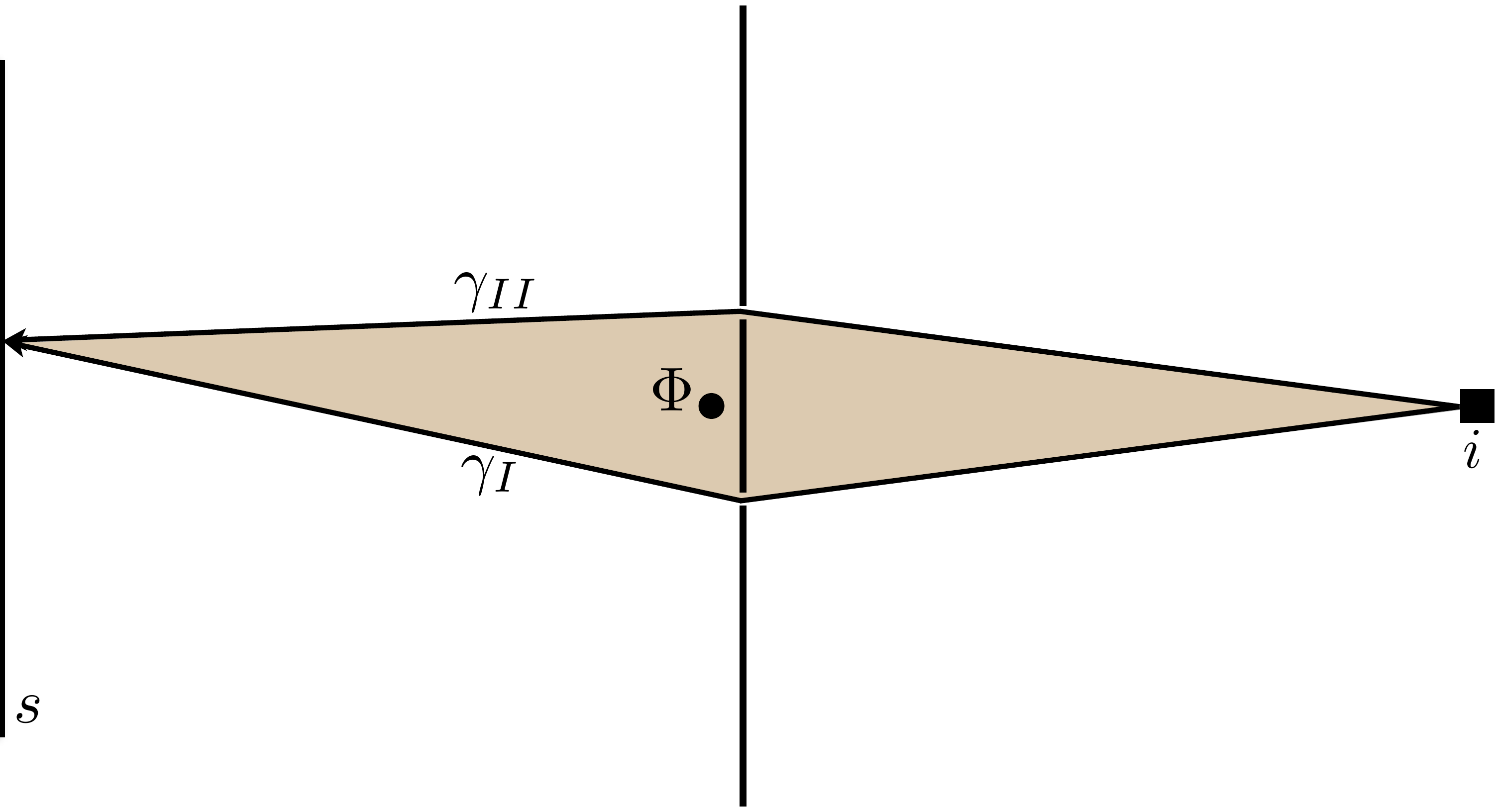}
  \caption{\label{fig:ABInterference} Elementary manifestation~\cite{AHARONOVBOHM,EHRENBERG,SAKURAI,FEYNMANII} of the Aharonov-Bohm effect in a double-slit thought experiment in which charged particles, emanating from the source $i$ to reach the screen $s$, pass either side of a thin solenoid placed behind the slit-system, between the two openings. The probability amplitudes associated with the two topologically distinct paths $\gamma_I$ and $\gamma_{II}$ acquire Dirac's magnetic phase factor, giving rise to a phase difference which amounts to the flux $\Phi$ enclosed by the curve which ensues after a reversal of either path.}
\end{figure}

An elementary realization of the effect, first put forward by Ehrenberg and Siday~\cite{EHRENBERG}, which also appears in the publication by Aharonov and Bohm~\cite{AHARONOVBOHM}, is now described in several textbooks (see, e.g.,~\cite{SAKURAI,FEYNMANII}). 
The thought experiment relies on a semiclassically inspired description in which the particle wave reaches a detector via mutually exclusive classical alternatives, $\gamma_I$ and $\gamma_{II}$ in figure~\ref{fig:ABInterference}, whereupon each contribution, due to a classically inaccessible magnetic flux $\Phi$, acquires Dirac's magnetic phase factor~\cite{DIRACII}. The probability amplitude which results for the particle wave emanating from the source $i$ to reach a specific point on the screen $s$ is then given by
\begin{align}
\label{eq:AbElementary}
  K &= K^{\gamma_I}_0 \exp \left( \frac{i}{\hbar} \frac{e}{c} \int_{\gamma_I} \vec{A} \cdot d\vec{x} \right) + K^{\gamma_{II}}_0 \exp  \left( \frac{i}{\hbar} \frac{e}{c} \int_{\gamma_{II}} \vec{A} \cdot d\vec{x} \right) \nonumber \\
     &=  \exp \left( \frac{i}{\hbar} \frac{e}{c} \int_{\gamma_I} \vec{A} \cdot d\vec{x} \right) \left( K^{\gamma_I}_0 + K^{\gamma_{II}}_0 \exp  \left( \frac{i}{\hbar} \frac{e}{c} \Phi \right) \right),
\end{align} 
where $K_0^{\gamma_{I/II}}$ denotes the contribution from the first/second path in the field-free case, and $\vec{A}$ is the vector potential 
giving rise to the magnetic flux $\Phi$. Taking Dirac's magnetic phase acquired along the first path out of the bracket in the second line in (\ref{eq:AbElementary}) corresponds to a reversal of this path in figure~\ref{fig:ABInterference}. Then the vector potential which comes with the second term in the second line in (\ref{eq:AbElementary}) is integrated along a contour enclosing the solenoid such that invoking Stoke's theorem produces a phase proportional to the magnetic flux $\Phi$. Thus the interference fringes on the screen are periodically shifted as a function of $\Phi$, while the envelope of the pattern remains invariant, which is confirmed by experiment~\cite{CHAMBERS,MOELLENSTEDTI,MOELLENSTEDTII}. None of the sources~\cite{AHARONOVBOHM,EHRENBERG,SAKURAI,FEYNMANII} of the above thought experiment, however, state how such a description is related to Schr\"odinger theory.

A precise relation assigning Dirac's magnetic phase to classical trajectories is established by the semiclassical approximation of the quantum propagator by Van Vleck and Gutzwiller~\cite{VANVLECK,MCGPI}
\begin{align}
\label{eq:SemicalssicalPropagator}
  \mathcal{K}\left( \vec{x}'',\vec{x}'; t'',t' \right) &= \frac{1}{\sqrt{2\pi i \hbar}^{f}} \sum_{\vec{x}_{cl}} \sqrt{\left| \det \left( -\frac{\partial^2 \mathcal{R}(\vec{x}'',\vec{x}';t'',t')}{\partial \vec{x}' \partial \vec{x}''} \right)\right| } \nonumber \\
  &\qquad \qquad \qquad \qquad \times \exp{\left( \frac{i}{\hbar} \mathcal{R}(\vec{x}'',\vec{x}';t'',t') -i \mu\frac{\pi}{2} \right)},
\end{align}
which constitutes an approximation of the probability amplitude of a particle starting out at $\vec{x}(t') = \vec{x}'$ to be later found at $\vec{x}(t'') = \vec{x}''$. Here $f$ denotes the number of degrees of freedom, and the sum runs over all classical trajectories $\vec{x}_{cl}$ which connect $\vec{x}'$ and $\vec{x}''$, entering Hamilton's principal function
\begin{align}
\label{eq:HPF1}
  \mathcal{R}(\vec{x}'',\vec{x}';t'',t') = \int_{t'}^{t''} \mathcal{L} \left( \vec{x}(t), \dot{ \vec{x}}(t), t \right) dt,
\end{align}  
where $\mathcal{L}$ is the Lagrangian of the system (for the origin and significance of the additional phase $\mu$ cf.~reference~\cite{MCGPI}). Thus each classical trajectory in (\ref{eq:SemicalssicalPropagator}) contributes a phase determined by Hamilton's principle function $\mathcal{R}$ in units of Planck's constant. If $\mathcal{L}$ contains a vector potential $\vec{A}$, the contribution of the latter constitutes precisely Dirac's magnetic phase factor.

\begin{figure}
\centering
  \subfloat[Process enclosing the flux\label{fig:ABTwoTrajectories1}]{%
    \includegraphics[width=0.4\textwidth]{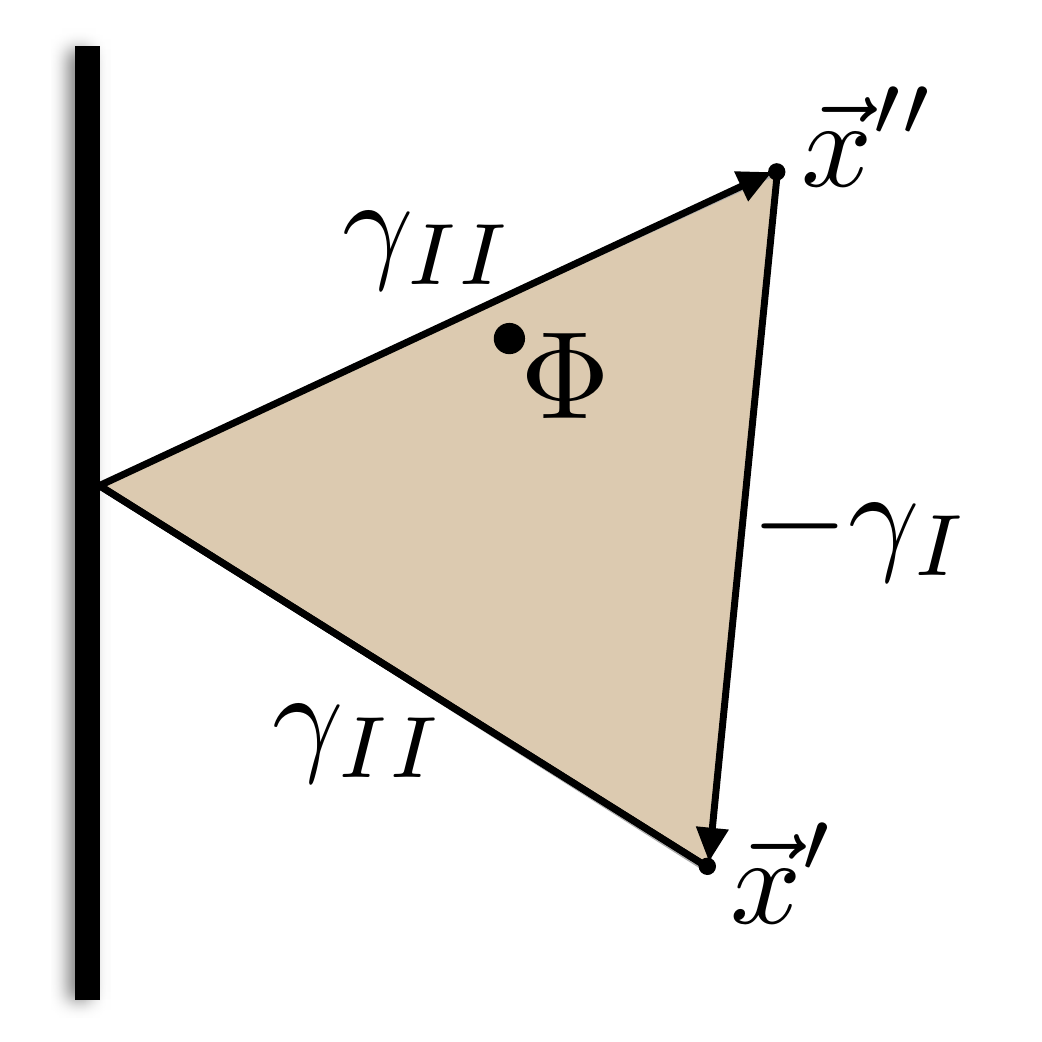}
  }
  \hfill
  \subfloat[Displacement of the endpoint\label{fig:ABTwoTrajectories2}]{%
    \includegraphics[width=0.4\textwidth]{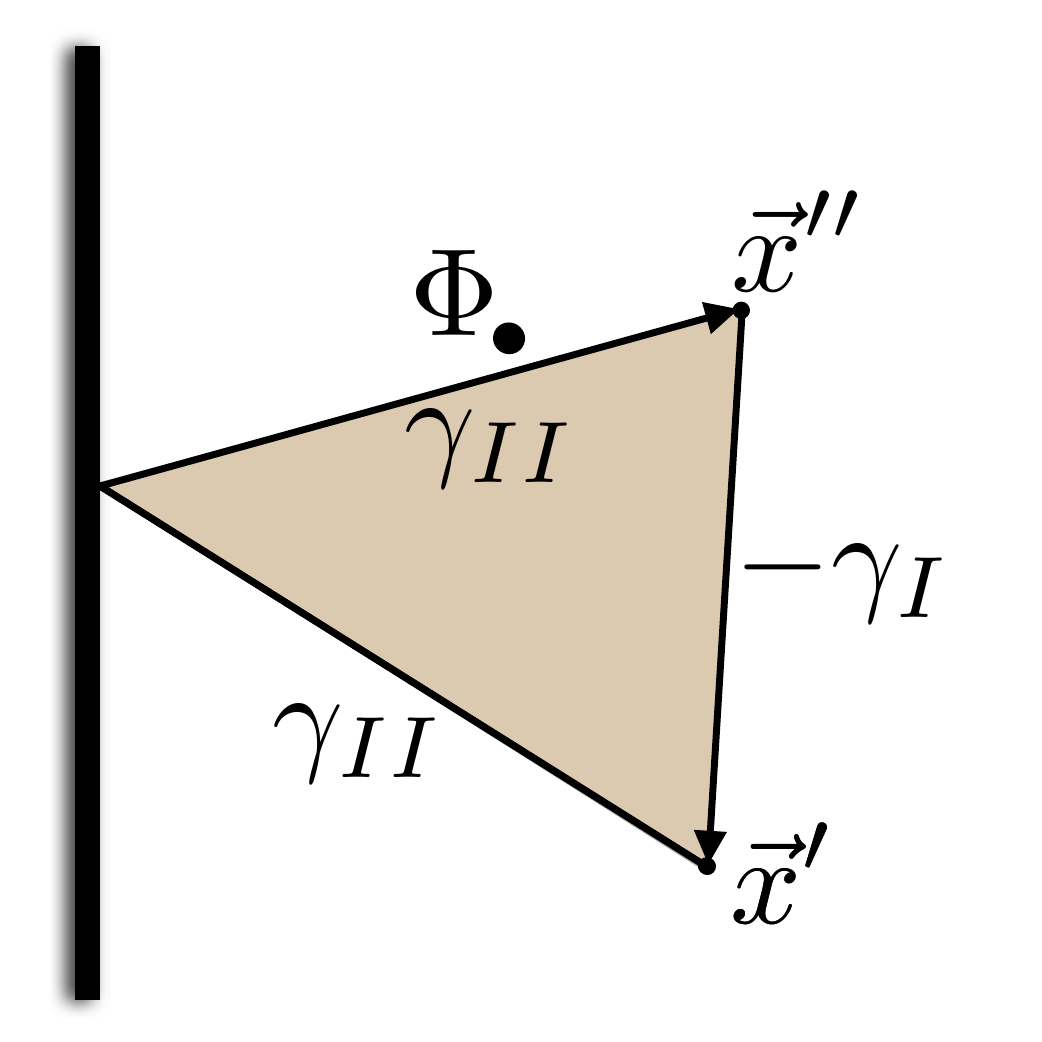}
  }
  
    \caption{Charged particle propagating from $\vec{x}'$ to $\vec{x}''$ in the presence of a magnetic string carrying magnetic flux $\Phi$, and in the vicinity of an infinitely extended wall. (a) As long the two possible trajectories, $\gamma_I$ unperturbed, and $\gamma_{II}$ once reflected at the wall, enclose the magnetic string, the semiclassical propagator~(\ref{eq:SemicalssicalPropagator}) generates a probability amplitude similar to~(\ref{eq:AbElementary}) as encountered in the elementary manifestation of the Aharonov-Bohm effect. (b) If the endpoint $\vec{x}''$ is displaced such that the closed curve ensuing after the reversal of either path no longer contains the magnetic string, the amplitude~(\ref{eq:AbElementary}) abruptly changes into~(\ref{eq:AbElementary2}) which is no longer a function of the flux $\Phi$. This renders~(\ref{eq:AbElementary}) discontinuous whenever a classical trajectory passes directly through the magnetic string.   }
  \label{fig:ABTwoTrajectories}
\end{figure}

In their original publication Aharonov and Bohm introduced the vector potential~\cite{AHARONOVBOHM} 
\begin{align}
\label{eq:Aab}
   \vec{A} = A_r \vec{e}_r + A_\varphi \vec{e}_\varphi = \frac{\Phi}{2 \pi r} \vec{e}_\varphi,
\end{align}
of an infinitely long and infinitesimally thin solenoid, now called a magnetic string, which gives rise to a magnetic field confined along the $z$-axis. We restrict the discussion to the $x$-$y$-plane in which the magnetic string thus pierces the origin. The vector potential (\ref{eq:Aab}) generates the second term in the Lagrangian
\begin{align}
\label{eq:Lab}
  \mathcal{L}_{\chi} = \mathcal{L}_0 - \chi \dot{\varphi},
\end{align}
wherein $\mathcal{L}_0$ contains all remaining terms when no current runs in the solenoid, and where we have introduced the abbreviation
\begin{align}
\label{eq:magneticchi}
  \chi =  -\frac{e}{c} \frac{\Phi}{2\pi}.
\end{align}
The second term in (\ref{eq:Lab}) does not contribute to the Euler-Lagrange equations of motion, reflecting the fact that classically the particle does not feel any Lorentz force. Yet this term contributes to Hamilton's Principal function
\begin{align}
\label{eq:Rab}
  \mathcal{R}_\chi = \mathcal{R}_0 - \chi \int_{\varphi'}^{\varphi''} d\varphi
\end{align}
a magnetic phase factor which is proportional to the angle swept by the respective classical trajectory~\footnote{The authors would like to thank Michael V.~Berry for the helpful comment that this reflects the subtle classicality of the Aharonov-Bohm effect: While the corresponding flux is not present in Newtonian mechanics, it makes an appearance in Hamiltonian mechanics.   }.

\begin{figure}
\centering
  \subfloat[Process enclosing the flux\label{fig:ABTwoTrajectories3}]{%
    \includegraphics[width=0.4\textwidth]{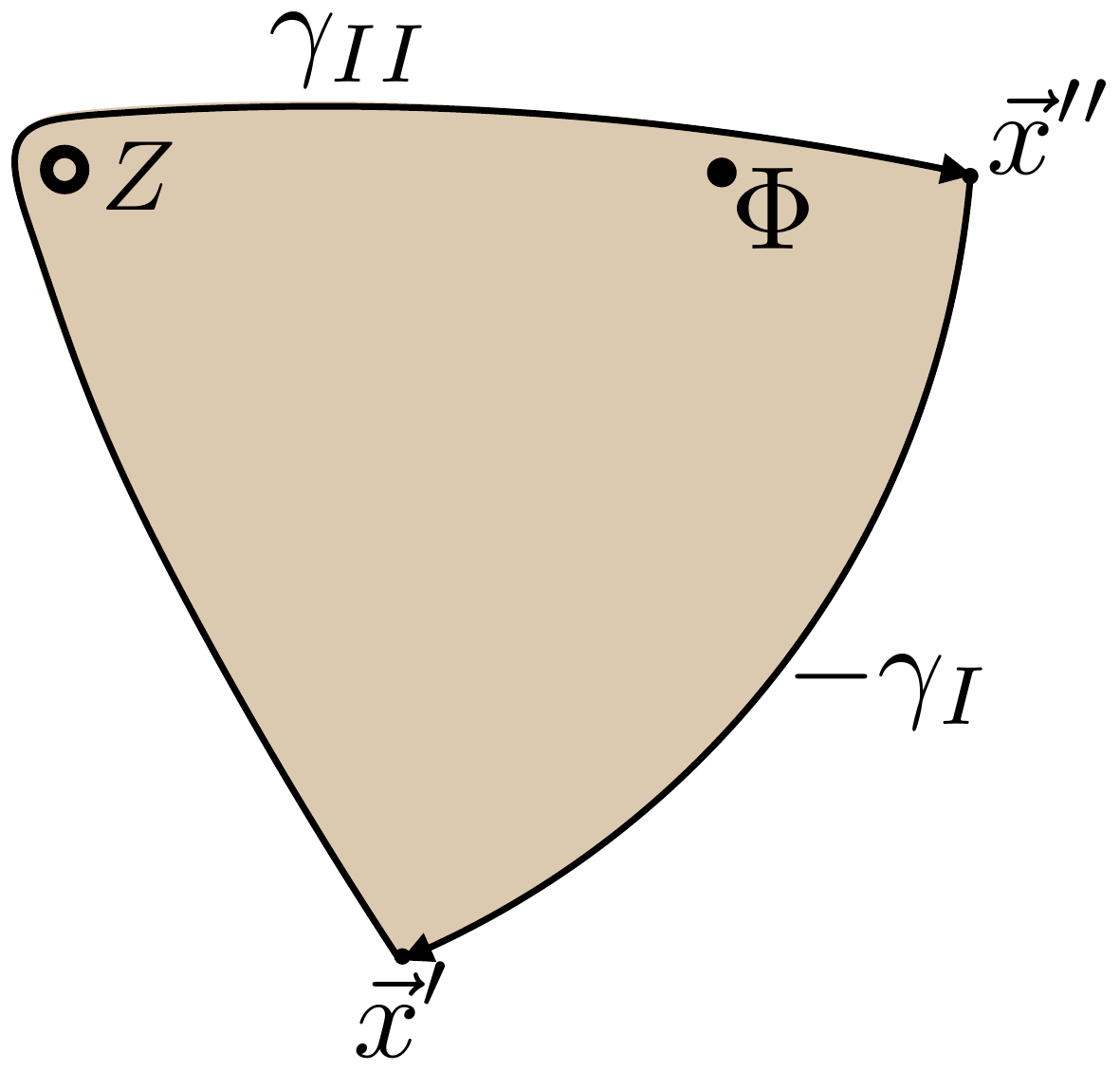}
  }
  \hfill
  \subfloat[Rotation of initial and endpoint\label{fig:ABTwoTrajectories4}]{%
    \includegraphics[width=0.4\textwidth]{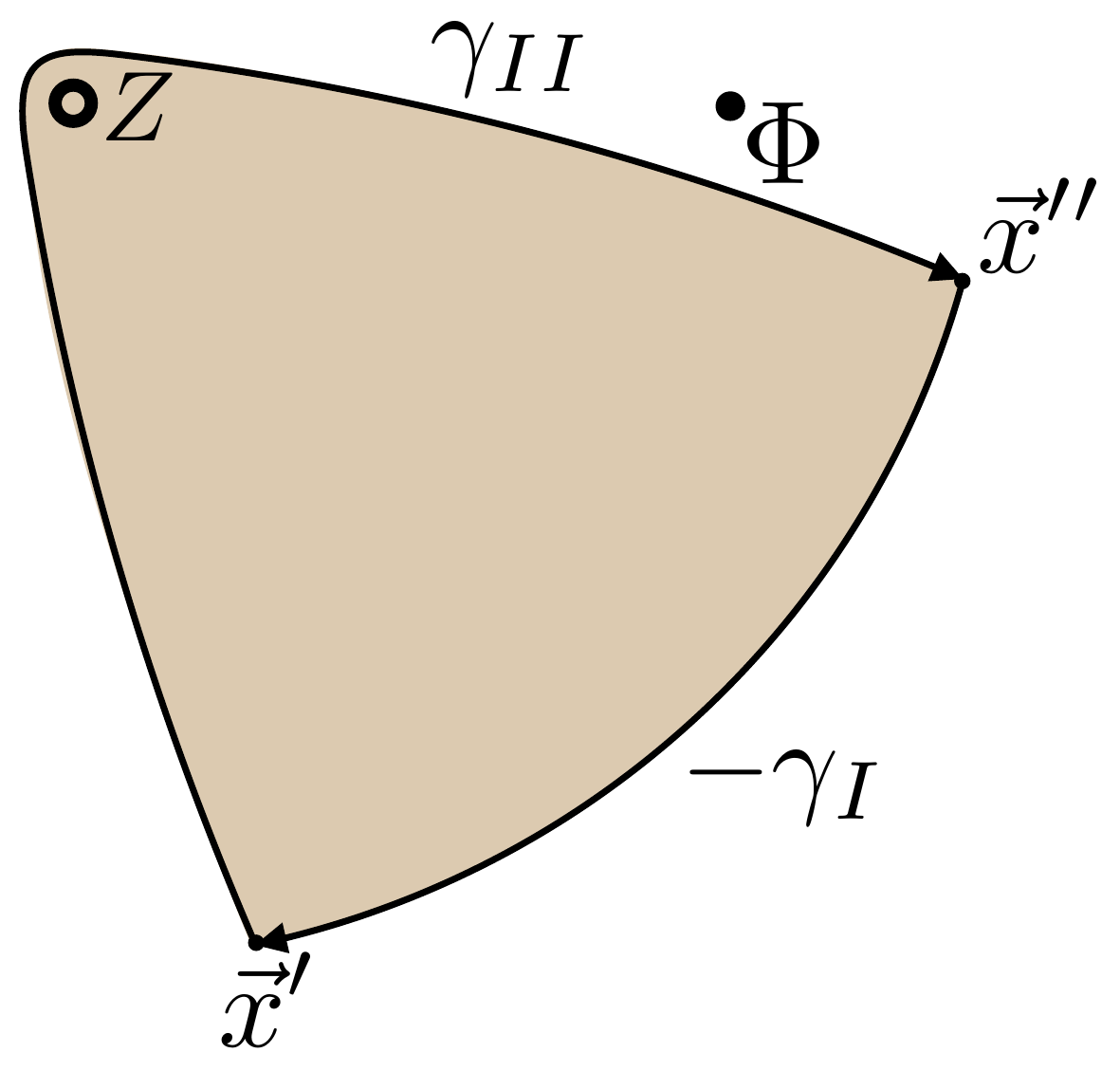}
  }  
    \caption{Charged particle propagating from $\vec{x}'$ to $\vec{x}''$, in the presence of an (infinitely massive) Coulomb attractor $Z$, as well as of a magnetic string carrying magnetic flux $\Phi$. The Coulomb potential gives rise to at least two trajectories~\cite{MCGPVIII} from $\vec{x}'$ to $\vec{x}''$, $\gamma_I$ connecting both points in a rather direct way, and $\gamma_{II}$ passing in a sharp turn behind the center of attraction. The classical trajectories are invariant under rotations of $\vec{x}'$ and $\vec{x}''$ about $Z$ such that the process in (a, adapted from~\cite{DISS}), enclosing the flux $\Phi$, can be transformed into the process depicted in (b), in which the trajectories no longer run around the magnetic string. Thus also in this scenario the semiclassical propagator~(\ref{eq:SemicalssicalPropagator}) features discontinuities whenever a classical trajectory passes directly through the flux. }
  \label{fig:ABTwoTrajectoriesSmooth}
\end{figure}

If the physical process is such that there exist exactly two trajectories starting at $\vec{x}'$ and ending at $\vec{x}''$, which enclose the magnetic string, insertion of (\ref{eq:Rab}) into (\ref{eq:SemicalssicalPropagator}) directly generates a probability amplitude similar to~(\ref{eq:AbElementary}). A conceivable situation which gives rise to such a process is depicted in figure~\ref{fig:ABTwoTrajectories1} where in addition to the flux the particle propagates in the presence of a wall that extends infinitely in the two-dimensional plane. Another possibility~\cite{DISS}, where the classical trajectories which enclose the flux are furthermore differentiable, arises if the particle in addition to the magnetic string propagates under the influence of a Coulomb attractor $Z$. This also gives rise to at least two classical trajectories~\cite{MCGPVIII}, as depicted in figure~\ref{fig:ABTwoTrajectories3}. Common to both situations is that the associated semiclassical propagator (\ref{eq:SemicalssicalPropagator}) is a discontinuous function of the dynamical variables $\vec{x}'$ and $\vec{x}''$. For the propagator associated with the wall, this can be seen by slightly displacing the endpoint of the propagation process such that the closed curve which ensues by following $-\gamma_{I}$ after $\gamma_{II}$ no longer contains the flux, as depicted in figure~\ref{fig:ABTwoTrajectories2}. For the particle in the Coulomb potential the same follows after a rotation of initial and endpoint of the process about the center of attraction as depicted in figure~\ref{fig:ABTwoTrajectories4}. Thus the probability amplitude abruptly changes from~(\ref{eq:AbElementary}) to
\begin{align}
\label{eq:AbElementary2}
  K = \exp \left( \frac{i}{\hbar} \frac{e}{c} \int_{\gamma_I} \vec{A} \cdot d\vec{x} \right) \left(K^{\gamma_I}_0 + K^{\gamma_{II}}_0 \right)
\end{align}
whenever a classical trajectory runs directly through the magnetic string. 

The principal intention of the present article is to elucidate the nature of such discontinuities which arise when the semiclassical propagator is employed to describe inaccessible magnetic flux. To this end, we will introduce a novel semiclassical limit to determine the semiclassical asymptotics of a particle that propagates otherwise freely in the presence of a magnetic string.  

The paper is structured as follows: In section~\ref{sec:II} we present established exact representations of the Aharonov-Bohm propagator, from which we derive our novel semiclassical limit in section~\ref{sec:III}. Section~\ref{sec:Discussion} contains a discussion of the interference process implied by the asymptotic limit. In section~\ref{sec:Whirlingwave} we elaborate on the relation of our asymptotic expressions to the so called whirling-wave representation of the exact propagator, before we conclude in section~\ref{sec:VI}.

\section{\label{sec:II}Exact solutions and representations of the Aharonov-Bohm propagator}

If no further potentials act on our charged particle except for a magnetic string, the Hamilton operator associated with the vector potential (\ref{eq:Aab}) reads
\begin{align}
\label{eq:Hab}
  H_\chi = -\frac{\hbar^2}{2m} \left( \frac{\partial^2}{\partial r^2} + \frac{1}{r}\frac{\partial}{\partial r} + \frac{1}{r^2} \left( \frac{\partial}{\partial \varphi} + \frac{i}{\hbar} \chi \right)^2 \right),
\end{align}
which has the eigenfunctions
\begin{align}
\label{eq:AbEf}
  \psi_{kl}\left(r, \varphi \right) = \left( -i \right)^{|l+\chi/\hbar|} \frac{1}{\sqrt{2\pi}} J_{|l+\chi/\hbar|} \left( kr \right)e^{i l \varphi}, \quad l \in \mathbb{Z}, \, k \in \mathbb{R}_0^+,
\end{align}
where $J$ denote the Bessel functions of the first kind~\cite{WATSON}. 
From (\ref{eq:AbEf}) Aharonov and Bohm construct their wave function~\cite{AHARONOVBOHM}
\begin{align}
\label{eq:abwf}
\psi_\chi \left(r, \varphi \right) = \sum_{l=-\infty}^\infty \left( -i \right)^{|l+\chi/\hbar|} J_{|l+\chi/\hbar|} \left( kr \right)e^{i l \varphi},
\end{align}
which describes the scattering of a plane wave incident on the magnetic string from the positive $x$-direction. The eigenfunctions~(\ref{eq:AbEf}) give rise to the completeness relation~\cite{WATSON}
\begin{align}
\label{eq:crab}
  \frac{\delta \left( r'' - r' \right)}{\sqrt{r''r'}} &\delta \left( \varphi'' - \varphi' \right)= \frac{1}{2\pi} \sum_{l = -\infty}^{\infty} \int_{0}^{\infty} dk\,k\, J_{|l+\chi/\hbar|} \left( kr'' \right) J_{|l+\chi/\hbar|} \left( kr' \right) e^{i l \left(\varphi''-\varphi' \right)}.
\end{align}
With the help of the latter one obtains the quantum propagator~\cite{BERNIDOINOMATAII}
\begin{align}
\label{eq:prab}
 K_\chi &= \frac{1}{2\pi} \sum_{l = -\infty}^{\infty} \int_{0}^{\infty} dk\,k\,\exp\left({-\frac{i}{\hbar} \frac{\hbar^2 k^2}{2m} \left( t''-t' \right)}\right)  J_{|l+\chi/\hbar|} \left( k r'' \right) J_{|l+\chi/\hbar|} \left( k r' \right) e^{i l \left(\varphi''-\varphi' \right)} \nonumber \\
  &= \frac{m}{2\pi i  \hbar \left( t''-t' \right)} \exp{ \left( \frac{i}{\hbar} \frac{m}{2}\frac{r''^2 +r'^2}{t''-t'} \right) }\sum_{l=-\infty}^\infty I_{\left| l+\chi/\hbar \right|}\left( -\frac{i}{\hbar} \frac{mr'' r'}{t''-t' } \right) e^{i {l \left( \varphi''-\varphi' \right)}},
\end{align}
where $I$ denote the modified Bessel functions of the first kind~\cite{WATSON}. The Bessel functions of the first kind $J$ and the modified Bessel functions of the first kind $I$ are related by~\cite{WATSON}
\begin{align}
  \left( -i \right)^{m} J_{m}(x) = I_m (-ix),  
\end{align}
such that, after the identifications
\begin{align}
\label{eq:zphi}
 \varphi \sim \varphi''-\varphi', \quad kr \sim  z:= \frac{m}{\hbar} \frac{r'' r'}{t''-t' },
\end{align}
and apart from the overall factor in the second line in~(\ref{eq:prab}), 
the mathematical structure of the Aharonov-Bohm wave function~(\ref{eq:abwf}) and of the Aharonov-Bohm propagator~(\ref{eq:prab}) coincide. Therefore all following considerations apply to both objects interchangeably. There exists an integral representation of (\ref{eq:prab}), which reads~\cite{KRETZSCHMAR,OLARIUPOPESCU,SIEBER} 
\begin{align}
\label{eq:PropagatorIntegral}
  & K_\chi = K_0 \Bigg( \exp{ \left( - \frac{i}{\hbar} \chi \left( \varphi'' - \varphi' \right) \right)  }   
   - \frac{1}{2} \mathrm{e}^{i l_\chi \left( \varphi''- \varphi' \right) -i \frac{\pi}{2} \left( l_\chi + \chi/\hbar \right) }  \sin \left( \left( l_\chi + \chi/\hbar \right) \pi \right) \nonumber \\
   & \quad\times \int_z^{\infty} \mathrm{e}^{i \zeta \cos \left( \varphi''- \varphi' \right) } \left( H_{1 - l_\chi - \chi/\hbar}^{(1)} \left( \zeta \right) \right.  \left. - i \mathrm{e}^{-i\left( \varphi''- \varphi' \right) } H_{- l_\chi - \chi/\hbar}^{(1)} \left( \zeta \right)  \right) d\zeta \Bigg),
\end{align}
and which will serve for later numerical comparisons. 
Here $H^{(1)}$ denote the Hankel functions of the first kind~\cite{WATSON}, and for the validity of~(\ref{eq:PropagatorIntegral}) the angular difference in the first term must be constrained to values $|\varphi''- \varphi'| < \pi$. The integer $l_\chi$ is determined by
\begin{align}
\label{eq:lchi}
  0 \leq l_\chi + \chi/\hbar < 1, 
\end{align}
and $K_0$ denotes the quantum propagator of the free particle
\begin{align}
\label{eq:freeK}
  K_0 = \frac{m}{2\pi i  \hbar \left( t''-t' \right)} \exp{ \left( \frac{i}{\hbar} \frac{m}{2}\frac{\left(\vec{x}'' - \vec{x}'\right)^2}{t''-t'} \right) }.
\end{align}

For half-integer flux $\chi/\hbar$, the propagator~(\ref{eq:prab},\ref{eq:PropagatorIntegral}) can be written in closed form~\cite{AHARONOVBOHM}, which we state here as
\begin{align}
\label{eq:semiflux}
  & K_\chi = K_0 \bigg( \exp\left(-\frac{i}{\hbar} \chi \left(\varphi''- \varphi' \right)  \right)  \frac{1}{\sqrt{i\pi}} \int_{-\sqrt{2z}\cos((\varphi''-\varphi')/2)}^\infty dx \, \exp \left( i x^2 \right)  \nonumber \\
    + &\exp\left( -\frac{i}{\hbar} \chi \left(\varphi''- \varphi' \mp 2\pi \right) \right) \frac{1}{\sqrt{i\pi}} \int_{-\infty}^{-\sqrt{2z}\cos((\varphi''-\varphi')/2)} dx \, \exp \left( i x^2 \right) \bigg),
\end{align}
where the sign in the exponential prefactor of the second line is given by that of $\varphi''-\varphi'$.

\section{\label{sec:III}Semiclassical approximation and the asymptotic limit}

Since the magnetic string does not influence the classical motion of the otherwise free particle, only one straight line trajectory classically connects $\vec{x}'$ and $\vec{x}''$ in the particle-flux line system described by~(\ref{eq:Hab}).
Hamilton's principal function~(\ref{eq:Rab}) of this classical trajectory, which we for later purposes express in polar coordinates, reads
\begin{align}
\label{eq:HPF}
  \mathcal{R}_\chi = \frac{m}{2} \frac{r''^2 + r'^2 - 2r'' r' \cos \left( \varphi'' - \varphi'\right)}{ t'' -t' } - \chi \int_{\varphi'}^{\varphi''} d\varphi,
\end{align}
where the first term is the free-particle contribution, and the second term represents the contribution of the magnetic string giving rise to the vector potential~(\ref{eq:Aab}). 
The canonical angular momentum, which is a conserved quantity along the classical trajectory, will play an important role in our following analysis and is given by
\begin{align}
\label{eq:clcanangmom}
\mathfrak{p}_{\varphi} = m r'' r' \frac{ \sin \left( \varphi'' - \varphi' \right)}{  t''-t' } - \chi.
\end{align}
Insertion of (\ref{eq:HPF}) into (\ref{eq:SemicalssicalPropagator}) produces the semiclassical approximation
\begin{align}
\label{eq:absemprop}
  \mathcal{K}_\chi = \mathcal{K}_0 \exp{\left( - \frac{i}{\hbar} \chi \int_{\varphi'}^{\varphi''}d\varphi \right)},
\end{align}
where the semiclassical propagator of the free particle $\mathcal{K}_0$ coincides with the exact expression~(\ref{eq:freeK}). Due to the restriction imposed on the angular difference which appears in the first term of (\ref{eq:PropagatorIntegral}), this first term coincides with (\ref{eq:absemprop}). The latter features a discontinuity whenever the classical trajectory associated with the propagation process directly passes through the magnetic string (in~(\ref{eq:PropagatorIntegral}) this discontinuity is compensated by the second term~\cite{SIEBER}), exactly as in our example systems of section~\ref{sec:intro}. Our main purpose here is to elucidate the origin of these discontinuities. To this end, we introduce a novel semiclassical limit which transforms the full solution~(\ref{eq:prab}) for the otherwise free particle into the semiclassical approximation~(\ref{eq:absemprop}). 

\begin{figure}
\centering
  \includegraphics[width=0.5\textwidth]{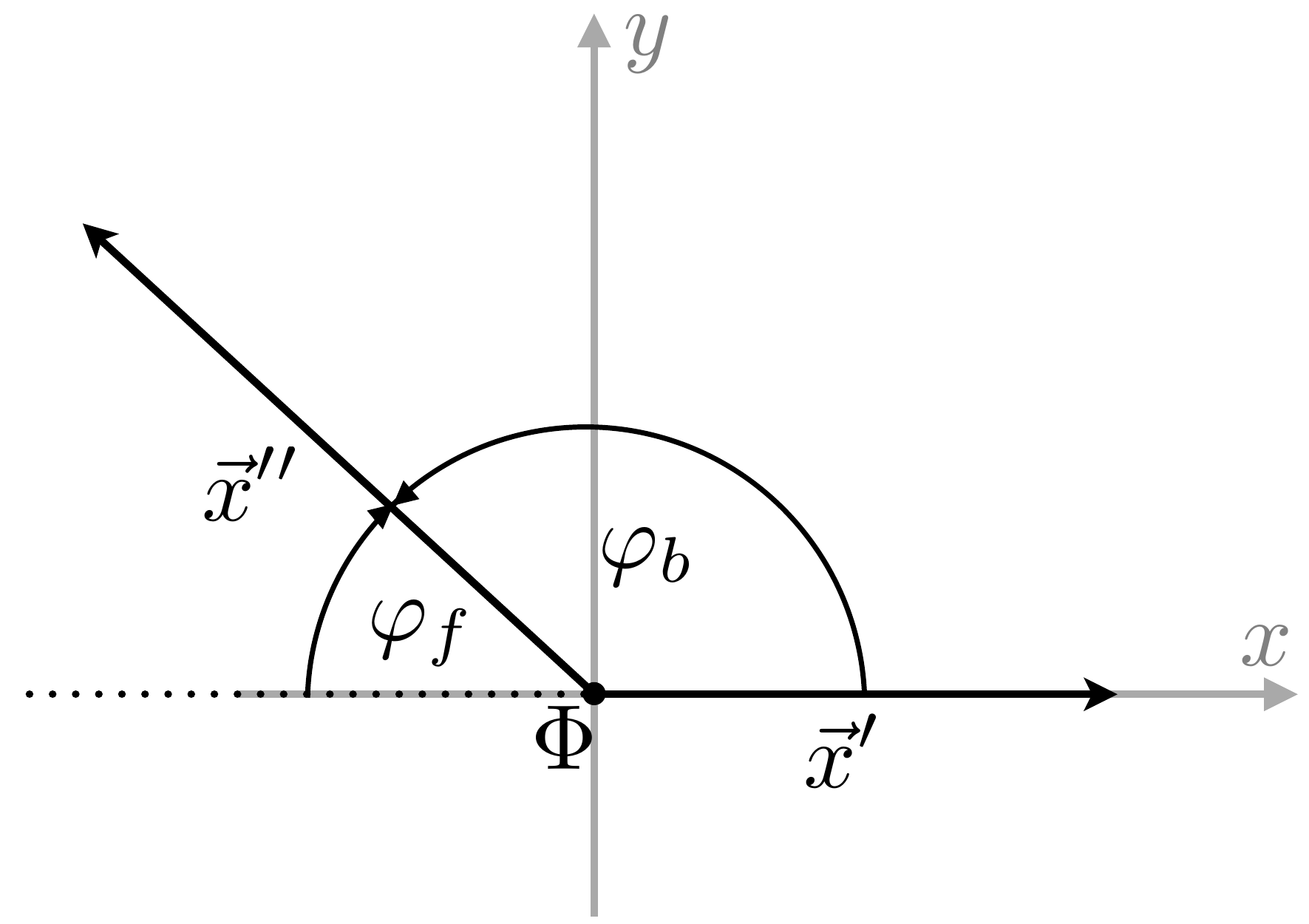}
  \caption{\label{fig:Sketch}(Adapted from~\cite{DISS}) Coordinate system in which the magnetic string, carrying the flux $\Phi$, pierces the origin of the two-dimensional plane, and $\vec{x}'$ is aligned along the positive $x$-axis. 
 For processes from $\vec{x}'$ to $\vec{x}''$, small angles around $\varphi_b=0$ define the backward direction, and the same applies for $\varphi_f=0$ 
  for processes in forward direction.}
\end{figure}
The procedure~\cite{DISS} is based on the exact asymptotic expansion~\cite{GRADSHTEYNRYZHIK} 
\begin{align}
 \label{eq:BesselExpand}
  &I_\nu(x) \sim \frac{e^x}{\sqrt{2\pi x}}\sum_{k =0}^{\infty} \frac{(-1)^k}{(2x)^k} \frac{\Gamma(\nu + k + \frac{1}{2})}{k! \Gamma(\nu - k + \frac{1}{2})} \nonumber \\
  &+\frac{\exp\left( -x - \left( \nu + \frac{1}{2} \right) \pi i \right)}{\sqrt{2\pi x}} \sum_{k =0}^{\infty} \frac{1}{(2x)^k} \frac{\Gamma(\nu + k + \frac{1}{2})}{k! \Gamma(\nu - k + \frac{1}{2})}
\end{align}
for the modified Bessel functions $I$ in~(\ref{eq:prab}), for large arguments $x$, valid in the range $-3\pi/2 < \arg(x) < \pi/2$. In our limit, the expansion~(\ref{eq:BesselExpand}) turns into the asymptotic approximation~\footnote{This corresponds to the usage of the Debye formula~\cite{ABRAMOWITZ} employed in reference~\cite{BERRYVIII}.}
\begin{align}
\label{eq:modbesselexpandlimit}
  I_\nu(x) \simeq \frac{1}{\sqrt{2\pi x}} \left( \exp \left( x -\frac{\nu^2}{2x} \right) +  \exp\left( -x - \left( \nu + \frac{1}{2} \right) \pi i +\frac{\nu^2}{2x} \right)  \right)
\end{align} 
as we will see in the following. For this purpose, we rewrite the sum in the exact propagator~(\ref{eq:prab}) as
\begin{align}
\label{eq:SuggestiveSum}
  \sum_{l=-\infty}^\infty I_{\left| l+\chi/\hbar \right|}\left(  \frac{\hbar}{i} \frac{mr'' r'}{ t''-t'} \frac{1}{\hbar^2} \right) e^{\frac{i}{\hbar} {\hbar l \frac{\varphi_b}{t''-t'}}\left( t''-t' \right)},
\end{align}
where 
\begin{align}
\label{eq:varphib}
  \varphi_b := \left( \varphi'' - \varphi' \right) \mathrm{mod}\, 2\pi \quad \in (-\pi,\pi),
\end{align}
see figure~\ref{fig:Sketch}, such that eventual excess integer multiples of $2\pi$ in the angular difference appearing in the exponent in~(\ref{eq:SuggestiveSum}) are evaluated beforehand. The intention is now to send $\hbar$, as it appears in the index and in the denominator of the argument of the modified Bessel functions in~(\ref{eq:SuggestiveSum}), to zero in the asymptotic expansion~(\ref{eq:BesselExpand}) with $\nu = \left| l+\chi/\hbar \right|$ and $x = -iz$ as defined in~(\ref{eq:zphi}), except for the exponentials in front of the infinite sums therein. Also in the numerator of the exponential of~(\ref{eq:SuggestiveSum}) we let $\hbar$ go to zero. 
The prefactor $1/ \sqrt{2\pi x}$ in~(\ref{eq:BesselExpand}) turns into
\begin{align}
  \frac{1}{\sqrt{2\pi x}} = \sqrt{ \frac{i}{\hbar} \frac{ t}{\pi\kappa} }\left(  \hbar\left( l + 1 \right) - \hbar l \right),
\end{align}
where we have introduced the abreviations
\begin{align}
\kappa = 2m r'' r' \quad \mathrm{and} \quad t = t''-t'.
\end{align}
Thus, following the above steps, the product
\begin{align}
\label{eq:continuousp}
  p_\varphi = \hbar l
\end{align}
turns into a continuous variable, and the sum over the quantum mechanically allowed values $l$ of the canonical angular momentum in~(\ref{eq:SuggestiveSum}) turns into an integral over $p_\varphi$ as given by~(\ref{eq:continuousp}), 
\begin{align}
\label{eq:sumtoint}
  \sum_{l=-\infty}^\infty \hbar\left( l + 1 \right) - \hbar l \to \int_{-\infty}^{\infty} dp_\varphi.
\end{align}

In order to see now how~(\ref{eq:BesselExpand}) turns into~(\ref{eq:modbesselexpandlimit}), we write out the first few terms of the first sum in~(\ref{eq:BesselExpand}) 
\begin{align}
\label{eq:BesselSeries}
  \begin{aligned}
    &\sum_{k =0}^{\infty} \frac{(-1)^k}{(2x)^k} \frac{\Gamma(\nu + k + \frac{1}{2})}{k! \Gamma(\nu - k + \frac{1}{2})} &=      & \frac{1}{0!}\frac{(-1)^0}{(2x)^0} \bigg(                             &  &                                        &&                                   &   &\nu^0      \bigg)   \nonumber \\
    & &                                                      +& \frac{1}{1!}\frac{(-1)^1}{(2x)^1} \bigg(                             &  &                                        &-\frac{1}{4}     &           & +& \nu^2  \bigg)   \nonumber \\
    & &                                                      +& \frac{1}{2!}\frac{(-1)^2}{(2x)^2} \bigg(                             &    \frac{9}{16}&                  &-\frac{5}{2}     &\nu^2 & +& \nu^4 \bigg)  \nonumber \\
    & &                                                      +& \frac{1}{3!}\frac{(-1)^3}{(2x)^3} \bigg( -\frac{225}{64}  & +\frac{259}{16}& \nu^2   &-\frac{35}{4}   &\nu^4 &+ &\nu^6 \bigg)  +  \dots 
  \end{aligned}\\
\end{align}
which continues in this manner. Except for the alternating sign, the second series in (\ref{eq:BesselExpand}) is of the same structure. 
Since the last column in the brackets in~(\ref{eq:BesselSeries}) coincides with the power series of an exponential, we find that~(\ref{eq:BesselExpand}) reduces to~(\ref{eq:modbesselexpandlimit}) if all contributions to (\ref{eq:BesselSeries}) but this last column were to vanish. Upon insertion of the argument $x = \kappa/ 2 i t \hbar$, and of the index $\nu = \left| l+\chi/\hbar \right|$ of the modified Bessel functions in~(\ref{eq:SuggestiveSum}) into~(\ref{eq:BesselSeries}), we find
\begin{align}
\label{eq:BesselSeries2}
  \begin{aligned}
    & &      +& \left(\frac{i}{\hbar} \frac{t}{\kappa} \right)^0 \bigg(                             &  &                                        &&                                   &   & (p_\varphi+\chi)^0  \bigg)   \nonumber \\
    & &                                                      -& \left(\frac{i}{\hbar} \frac{t}{\kappa} \right)^1 \bigg(                             &  &                                        &-\frac{\hbar^2}{4}     &           & +& (p_\varphi+\chi)^2  \bigg)   \nonumber \\
    & &                                                      +& \left(\frac{i}{\hbar} \frac{t}{\kappa} \right)^2 \bigg(                             &    \frac{9\hbar^4}{16}&                  &-\frac{5\hbar^2}{2}     &(p_\varphi+\chi)^2 & +& (p_\varphi+\chi)^4 \bigg)  \nonumber \\
    & &                                                      -& \left(\frac{i}{\hbar} \frac{t}{\kappa} \right)^4 \bigg( -\frac{225\hbar^6}{64}  & +\frac{259\hbar^4}{16}& (p_\varphi+\chi)^2   &-\frac{35\hbar^2}{4}   &(p_\varphi+\chi)^4 &+ &(p_\varphi+\chi)^6 \bigg)  +  \dots, 
  \end{aligned}\\
\end{align}
meaning that the columns prior to the last one are suppressed with respect to the latter at least by the order $\hbar^2$, and indeed vanish in the limit which we discussed in connection with~(\ref{eq:continuousp}) and~(\ref{eq:sumtoint}). Thus the contribution to~(\ref{eq:prab}) from the first series in~(\ref{eq:BesselExpand}) turns in this limit into 
\begin{align}
\label{eq:Kbackward0}
 K_\chi \simeq \frac{m}{2\pi i  \hbar \left( t''-t' \right)} &\exp{ \left( \frac{i}{\hbar} \frac{m}{2}\frac{\left(r'' -r'\right)^2}{t''-t'} \right) } \nonumber \\
  &\times \sqrt{\frac{i}{\hbar} \frac{t}{\pi\kappa}} \int_{-\infty}^\infty dp_\varphi \, \exp \left(- \frac{i}{\hbar} \left( \frac{t}{\kappa} \left( p_\varphi + \chi \right)^2 -p_\varphi \varphi_b \right) \right).
\end{align}

The two 
parts of (\ref{eq:BesselExpand}) 
essentially differ by the
extra factor $\exp\left( - \left( \nu + \frac{1}{2} \right) \pi i \right)$ which comes with the second sum.
Due to this factor the contribution of the second series in (\ref{eq:BesselExpand}) vanishes in our limit:
When splitting the sum in (\ref{eq:SuggestiveSum}) 
at $l_\chi$, determined by~(\ref{eq:lchi}),
one obtains two sums where in each the index of the modified Bessel functions enters through 
the extra factor which comes with the second series of (\ref{eq:BesselExpand}), and contributes a phase of $\pm \pi l$.
Except for their sign, 
the resulting 
contributions of even and odd $l$ 
coincide in the limit as outlined above, and cancel.

After quadratic completion and evaluation of the Fresnel integral in~(\ref{eq:Kbackward0}) we obtain 
\begin{align}
\label{eq:Kbackward}
  K_\chi \simeq \frac{m}{2\pi i  \hbar \left(t''-t' \right)} \exp \Bigg( \frac{i}{\hbar} \Bigg( \frac{m}{2} \frac{\left( r''-r' \right)^2}{t''-t'}   + \frac{m}{2}  r'' r' \frac{\varphi_b^2}{t''-t'}  -\chi \varphi_b \Bigg)  \Bigg).
\end{align}
Close to the backward direction, \textit{i.e.} for small values of $\varphi_b$, insertion of~(\ref{eq:varphib}) into~(\ref{eq:HPF}), followed by a truncation of the cosine therein after the second order, produces the phase of~(\ref{eq:Kbackward}). Therefore~(\ref{eq:Kbackward}) and~(\ref{eq:absemprop}) coincide in the region of validity of this expansion, where our limiting procedure thus directly leads to the semiclassical approximation of Van Vleck/Gutzwiller. Since the integral in~(\ref{eq:Kbackward0}) is quadratic in $p_\varphi$, also an evaluation by the method of stationary phase produces the exact result~(\ref{eq:Kbackward}). The stationary point is in this case given by 
\begin{align}
  p_{\varphi} = m r'' r' \frac{   \varphi_b }{  t''-t' } - \chi
\end{align}
which, close to the backward direction, coincides with~(\ref{eq:clcanangmom}) such that our method here shows how the main contribution to the integral in~(\ref{eq:Kbackward0}) results from those canonical angular momenta which are close to the classical value. 

In order to determine the asymptotic behavior of~(\ref{eq:prab}) close to the forward direction we repeat the above steps, but only after insertion of 
\begin{align}
\label{eq:varphif}
  \varphi_f=\pm\pi -\varphi_b \quad \in (-\pi,\pi),
\end{align}  
see figure~\ref{fig:Sketch}, where the sign coincides with the sign of $\varphi_b$, into~(\ref{eq:SuggestiveSum}).
Due to the extra term $\pm \pi$ in the definition~(\ref{eq:varphif}), which enters the exponential in~(\ref{eq:SuggestiveSum}) as a phase, the same limiting procedure as in the above derivation of (\ref{eq:Kbackward})
now suppresses the first series in (\ref{eq:BesselExpand}), while the second series renders
\begin{align}
\label{eq:Kforward0}
 K_\chi \simeq \frac{m}{2\pi i \hbar (t''-t')} &\exp{ \left( \frac{i}{\hbar} \frac{m}{2}\frac{\left(r'' + r'\right)^2}{t''-t'} \right) }  \nonumber \\
  &\times \sqrt{\frac{1}{i\hbar}\frac{t}{ \pi\kappa} } \left(  \int_{-\chi}^\infty dp_\varphi \, \exp \left( \frac{i}{\hbar} \left(  \frac{t}{\kappa} \left( p_\varphi + \chi \right)^2 -p_\varphi \varphi_f - \chi\pi \right) \right) \right. \nonumber \\
&\qquad\qquad\left. + \int_{-\infty}^{-\chi} dp_\varphi \, \exp \left( \frac{i}{\hbar} \left( \frac{t}{\kappa} \left( p_\varphi + \chi \right)^2 -p_\varphi \varphi_f + \chi\pi \right) \right) \right) .
\end{align}
The finite integration bounds in (\ref{eq:Kforward0}) 
are determined by multiplication of the constraints on $l_\chi$ stated in (\ref{eq:lchi}) by $\hbar$, and sending the latter to zero. 
Since the kinetic angular momentum 
is related to the canonical angular momentum 
by $p_\varphi^{\text{kin}} = p_\varphi + \chi$, 
in the above limit all positive kinetic angular momenta contribute to the first integral in (\ref{eq:Kforward0}), whereas the negative 
kinetic angular momenta contribute to the second. 

Due to the finite integration bounds, evaluation of~(\ref{eq:Kforward0}) via the method of stationary phase in this case constitutes an approximation. This produces
\begin{align}
\label{eq:Kforwardplus}
  K_\chi \simeq \frac{m}{2\pi i  \hbar \left(t''-t' \right)} \exp \Bigg( \frac{i}{\hbar} \Bigg( \frac{m}{2} \frac{\left( r''+r' \right)^2}{t''-t'}   - \frac{m}{2}  r'' r' \frac{\varphi_f^2}{t''-t'}  -\chi \left( \varphi_f - \pi \right) \Bigg)  \Bigg)
\end{align}
for $\varphi_f > 0$, and
\begin{align}
\label{eq:Kforwardminus}
  K_\chi \simeq \frac{m}{2\pi i  \hbar \left(t''-t' \right)} \exp \Bigg( \frac{i}{\hbar} \Bigg( \frac{m}{2} \frac{\left( r''+r' \right)^2}{t''-t'}   - \frac{m}{2}  r'' r' \frac{\varphi_f^2}{t''-t'}  +\chi \left( \varphi_f + \pi \right) \Bigg)  \Bigg)
\end{align}
for $\varphi_f < 0$. Close to the forward direction, both~(\ref{eq:Kforwardplus}) and~(\ref{eq:Kforwardminus}) coincide with~(\ref{eq:absemprop}) in the respective angular range, such that also in this case the connection to the semiclassical propagator of Gutzwiller is established. After the stationary phase approximation leading to~(\ref{eq:Kforwardplus}) and~(\ref{eq:Kforwardminus}), respectively, the continuity of~(\ref{eq:Kforward0}) is thus no longer resolved. The stationary point is in both cases given by
\begin{align}
  p_{\varphi} = m r'' r' \frac{   \varphi_f }{  t''-t' } - \chi
\end{align}
which also here, close to the forward direction, coincides with the classical value~(\ref{eq:clcanangmom}) of the canonical angular momentum.

To discuss in section~\ref{sec:Discussion} the interference process which smoothes out the discontinuity between~(\ref{eq:Kforwardplus},\ref{eq:Kforwardminus}) close to the forward direction, we render the integrals in~(\ref{eq:Kforward0}) dimensionless and perform a quadratic completion to cast the latter into the final form
\begin{align}
\label{eq:Kforward}
  K_\chi \sim \frac{m}{2\pi i  \hbar t} &\exp \Bigg( \frac{i}{\hbar} \Bigg( \frac{m}{2}\frac{\left(r'' + r'\right)^2 }{t''-t'} - \frac{m}{2}  r'' r' \frac{\varphi_f^2}{t''-t'}  \Bigg)  \Bigg) \nonumber \\ 
    &\times \Bigg( \exp \left( - \frac{i}{\hbar} \chi \left(\pi - \varphi_f \right)  \right) \frac{1}{\sqrt{i\pi}} \int_{-\sqrt{z/2} \varphi_f}^\infty dx \, \exp \left( i x^2 \right) \nonumber \\
&+ \exp \left( + \frac{i}{\hbar} \chi \left(\pi + \varphi_f\right)  \right)\frac{1}{\sqrt{i\pi}} \int_{-\infty}^{-\sqrt{z/2} \varphi_f} dx \, \exp \left( i x^2 \right) \Bigg).
\end{align}
Indeed, from our observation following (\ref{eq:Kforward0}) we find that the contributions of the positive/negative kinetic angular momenta acquire Dirac's magnetic phase as if encircling the solenoid not more than once in a counterclockwise/clockwise rotational sense.

Note that nowhere in our derivation of (\ref{eq:Kbackward},\ref{eq:Kforward}) did we explicitly require that $\varphi_b$ or $\varphi_f$ be close to zero. This requirement must be intricately related to our evaluation of the extra phases $\pm \pi l$, since this determines the range of validity of the expressions which result after $\hbar$ is sent to zero.

\section{\label{sec:Discussion}Interference process in the forward direction}

\begin{figure}
\centering
  \subfloat[$\Re (K_0^+/K_0)$\label{fig:ReKPlus}]{%
    \includegraphics[width=0.43\textwidth]{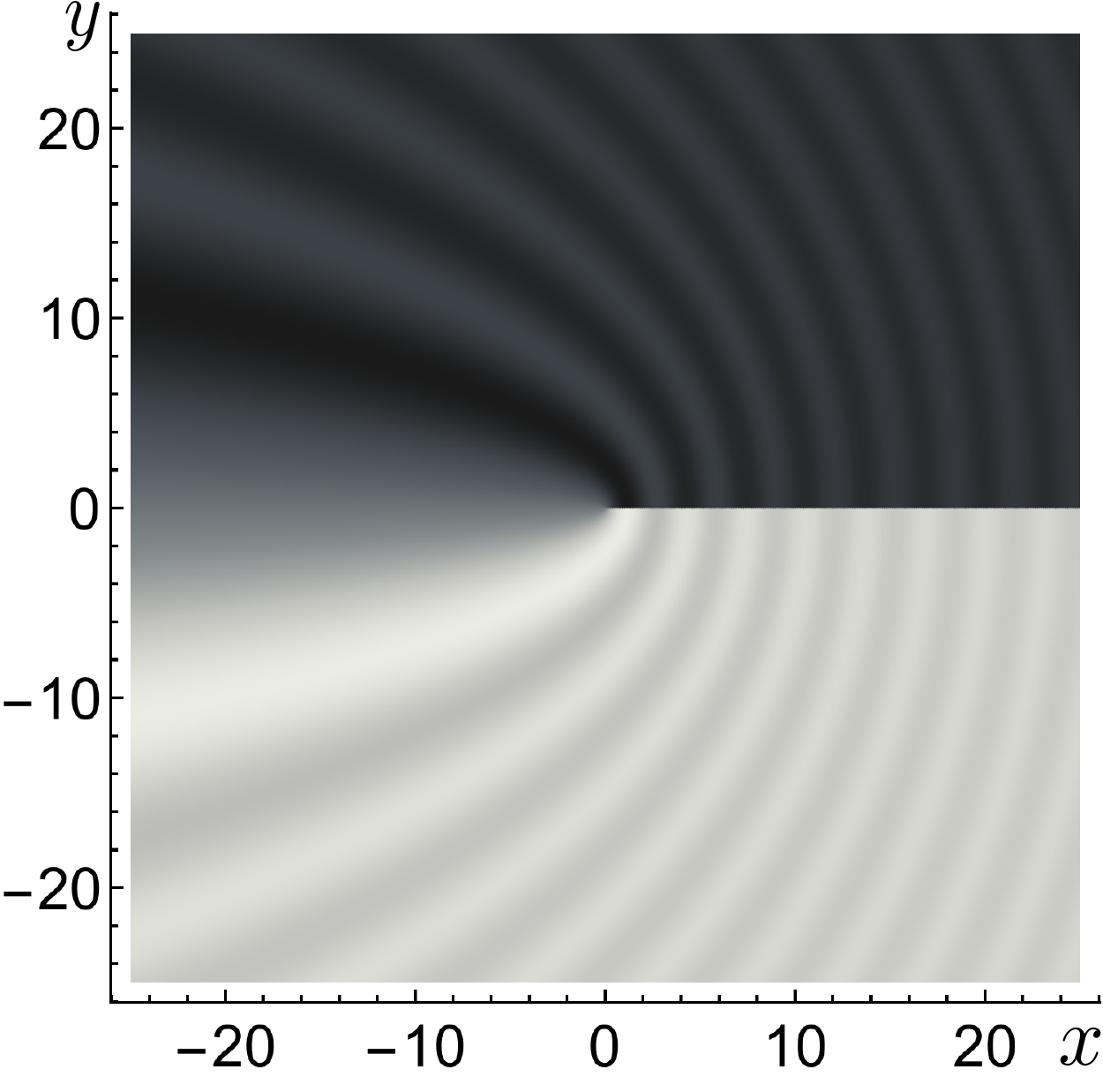}
  }
  \hfill
  \subfloat[$\Re (K_0^-/K_0)$\label{fig:ReKMinus}]{%
    \includegraphics[width=0.53\textwidth]{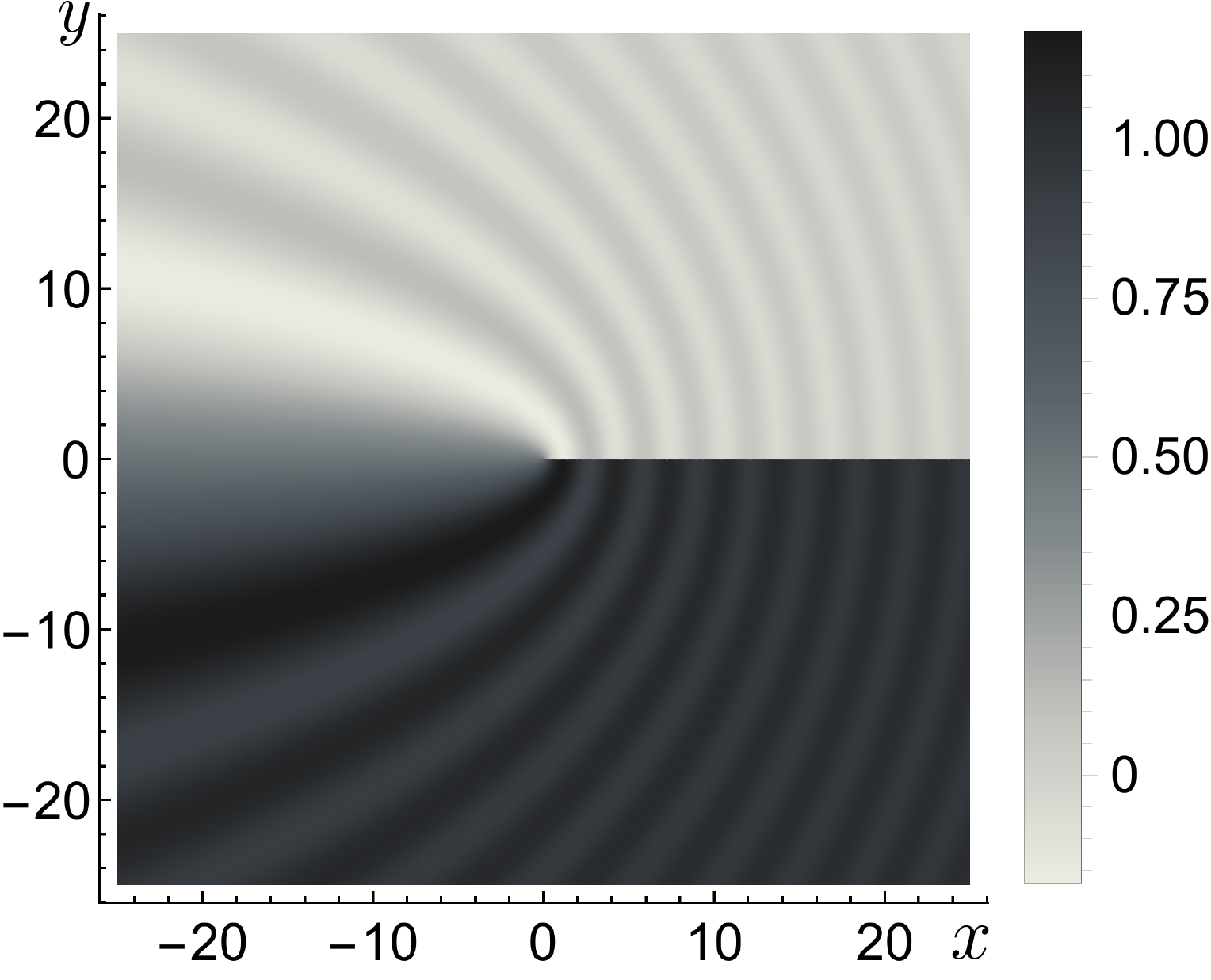}
  }  
    \caption{Real part of the half-waves $K_0^+$ and $K_0^-$ bending around the origin in a counterclockwise (a) and clockwise (b) rotational sense, respectively, divided by the free propagator factor $K_0$, where $x = z \cos \varphi_b$, $y = z \sin \varphi_b$. Amended accordingly by Dirac's magnetic phase, the superposition of these half-waves describes the forward Aharonov-Bohm interference process~(\ref{eq:Kforward},\ref{eq:KOP}) in the asymptotic limit. The sum $K_0^+ + K_0^-$ amounts to the free-particle propagator $K_0$.   }
  \label{fig:ReKPlusMinus}
\end{figure}

In a coordinate system in which $\vec{x}'$ coincides with the $x$-axis as in figure~\ref{fig:Sketch}, the free propagator~(\ref{eq:freeK}) can be split into two half-waves which on their way to $\vec{x}''$ in the forward direction pass the solenoid along the $y$-axis in an opposite rotational sense~\cite{OLARIUPOPESCU}. Thereby, the half-wave which passes the solenoid counterclockwise via the positive $y$-axis is given by~\cite{MORSEFESHBACH}\footnote{In reference~\cite{MORSEFESHBACH}, these half-waves originally appear in the context of knife-edge scattering.}
\begin{align}
\label{eq:KzeroPlus}
  K_0^+ = &K_0 \frac{1}{\sqrt{i\pi}} \int_{-\sqrt{2z}\sin(\varphi_f/2)}^{\infty} dx \, \exp \left( i x^2 \right)\, ,
\end{align} 
see figure~\ref{fig:ReKPlus}.
The half-wave passing the solenoid clockwise via the negative $y$-axis is related to (\ref{eq:KzeroPlus}) by
$K_0^- \left(\varphi_f \right) = K_0^+ \left(-\varphi_f \right)$, see figure~\ref{fig:ReKMinus}, and their sum $K_0^++K_0^-$ yields the free propagator~(\ref{eq:freeK}). The 
additional {\em ad-hoc} assumption that these contributions acquire magnetic phases as in (\ref{eq:absemprop}) then leads to the expression~\cite{OLARIUPOPESCU}\footnote{The expression originally given in reference~\cite{OLARIUPOPESCU} must be amended by the overall phase $\exp(i\chi\varphi_f/\hbar)$.}
\begin{align}
\label{eq:KOP}
  &K_\chi \sim K_0 \bigg( \exp\left(-\frac{i}{\hbar} \chi \left(\pi - \varphi_f \right)  \right)  \frac{1}{\sqrt{i\pi}} \int_{-\sqrt{2z}\sin(\varphi_f/2)}^\infty dx \, \exp \left( i x^2 \right)  \nonumber \\
    &\,\,\,+ \exp\left( +\frac{i}{\hbar} \chi \left(\pi + \varphi_f\right) \right) \frac{1}{\sqrt{i\pi}} \int_{-\infty}^{-\sqrt{2z}\sin(\varphi_f/2)} dx \, \exp \left( i x^2 \right) \bigg)
\end{align}
which was originally proposed for the forward direction, where it coincides with~(\ref{eq:Kforward}) from our novel derivation. The approximate propagator~(\ref{eq:KOP}) also appears as the lowest order~\footnote{Taking into account only the first term in equation~(20) of reference~\cite{BerryiX} establishes the connection to~(\ref{eq:KOP}). } of an asymptotic series expansion of~(\ref{eq:prab})~\cite{BerryiX}, where it is shown that the accuracy of~(\ref{eq:KOP}) increases in the entire angular range as $z$ increases.
(\ref{eq:KOP}), however, exhibits a discontinuity as $\varphi_f\rightarrow\pm\pi$, \textit{i.e.}~in backward direction, that vanishes only as $z\to\infty$. The discontinuity is, nonetheless, of the same order as the error of the approximation, and whether or not there exists an asymptotic approximation of~(\ref{eq:prab}) based on~(\ref{eq:KzeroPlus}) that is smooth in the entire angular range and containing only the contributions of maximally two topologically distinct paths is to date unknown~\cite{BERRYXI}. For half-integer flux (in units of $\hbar$), (\ref{eq:KOP}) coincides with the exact propagator~(\ref{eq:semiflux}). We will elaborate on this point in the next section. 

\begin{figure}
\centering
  \subfloat[$z = \pi$\label{fig:zh1}]{%
    \includegraphics[width=0.46\textwidth]{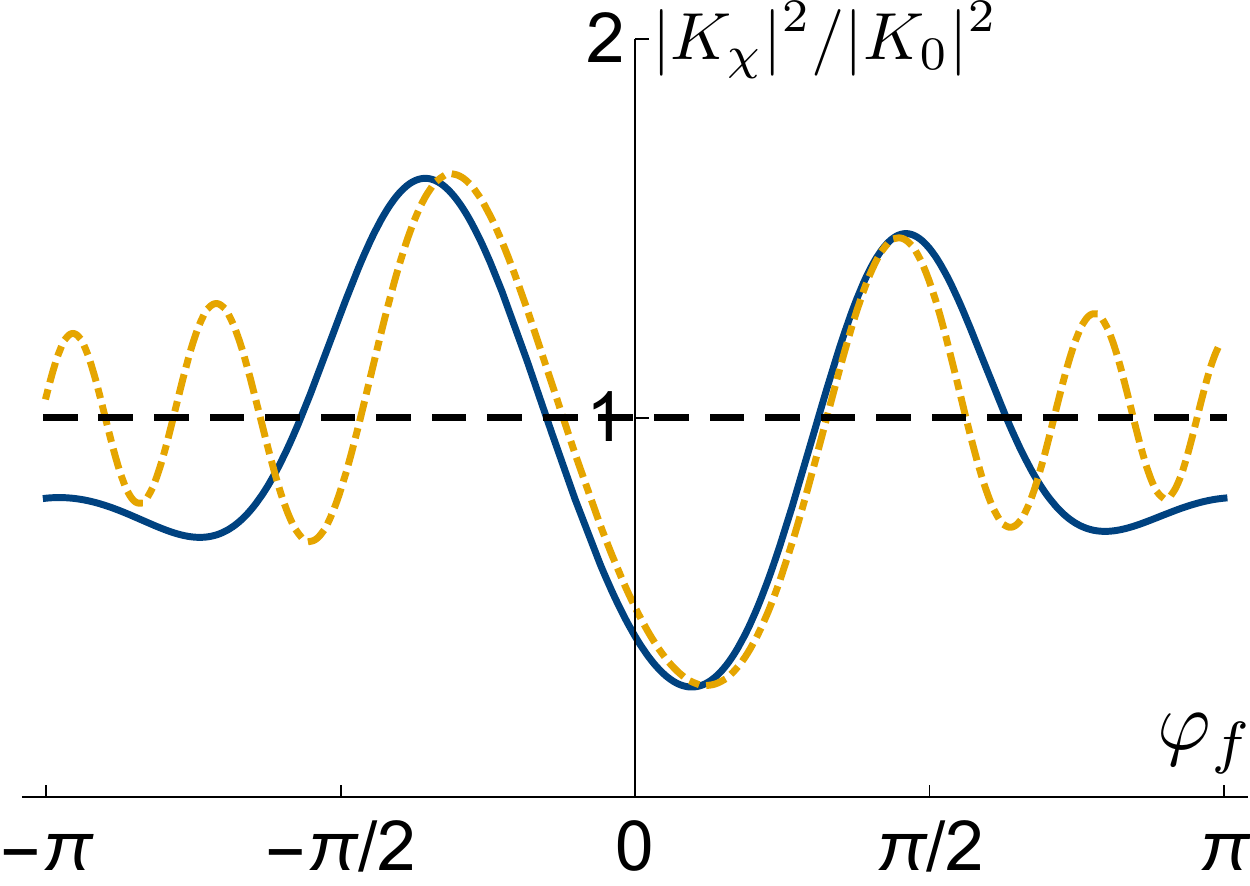}
  }
  \hfill
  \subfloat[$z = 3\pi$\label{fig:zh10}]{%
    \includegraphics[width=0.46\textwidth]{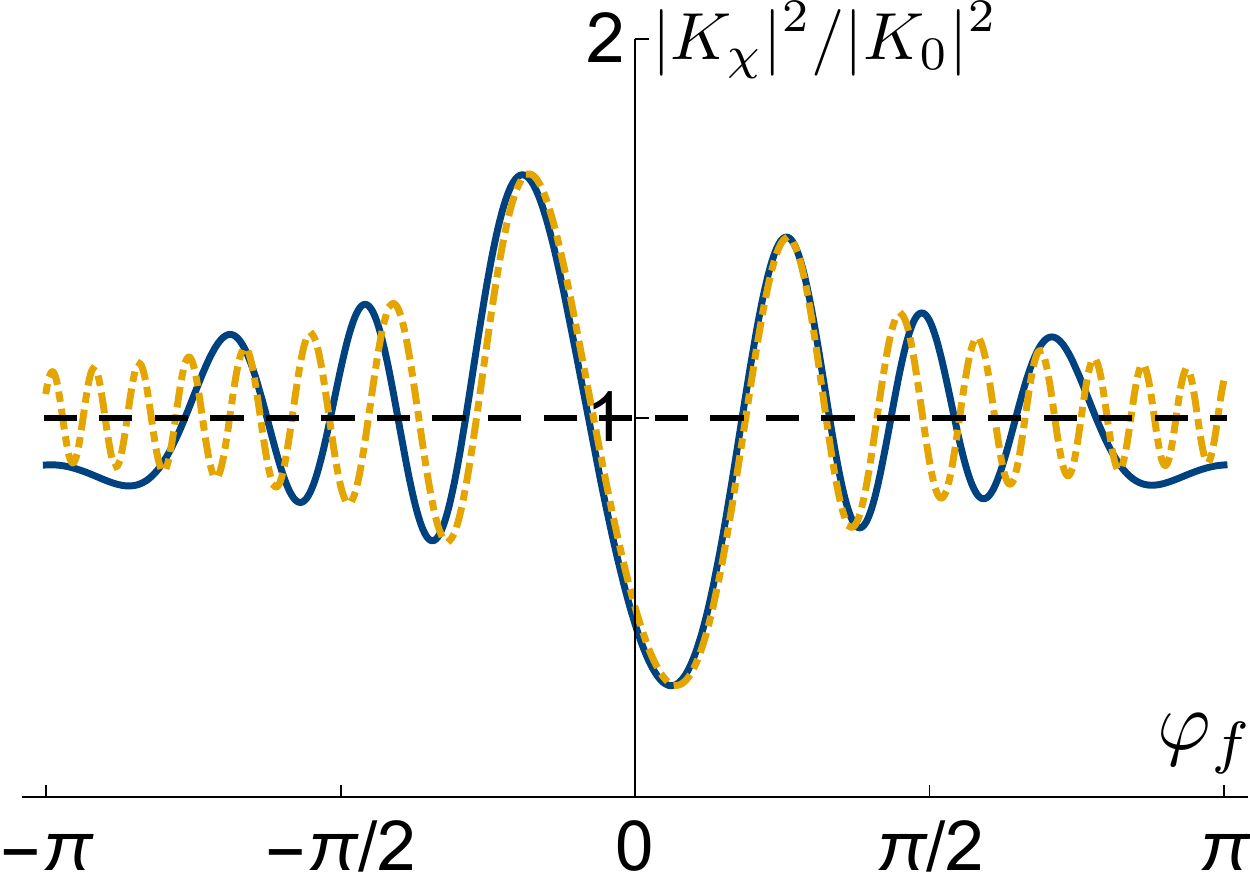}
  }
  
  \subfloat[$z = 10\pi$\label{fig:zh50}]{%
    \includegraphics[width=0.46\textwidth]{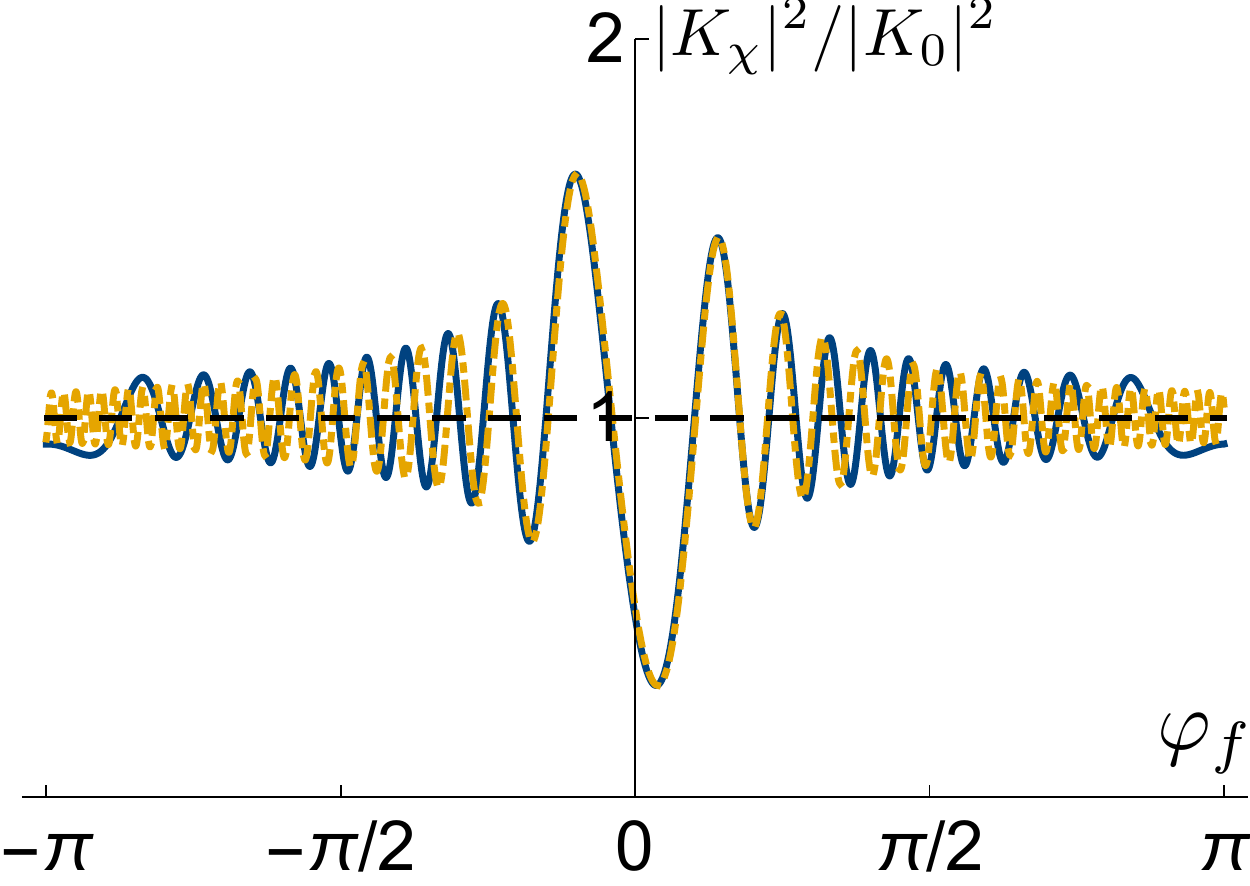}
  }
  \hfill
  \subfloat[$z = 10\pi$\label{fig:ph50}]{%
    \includegraphics[width=0.46\textwidth]{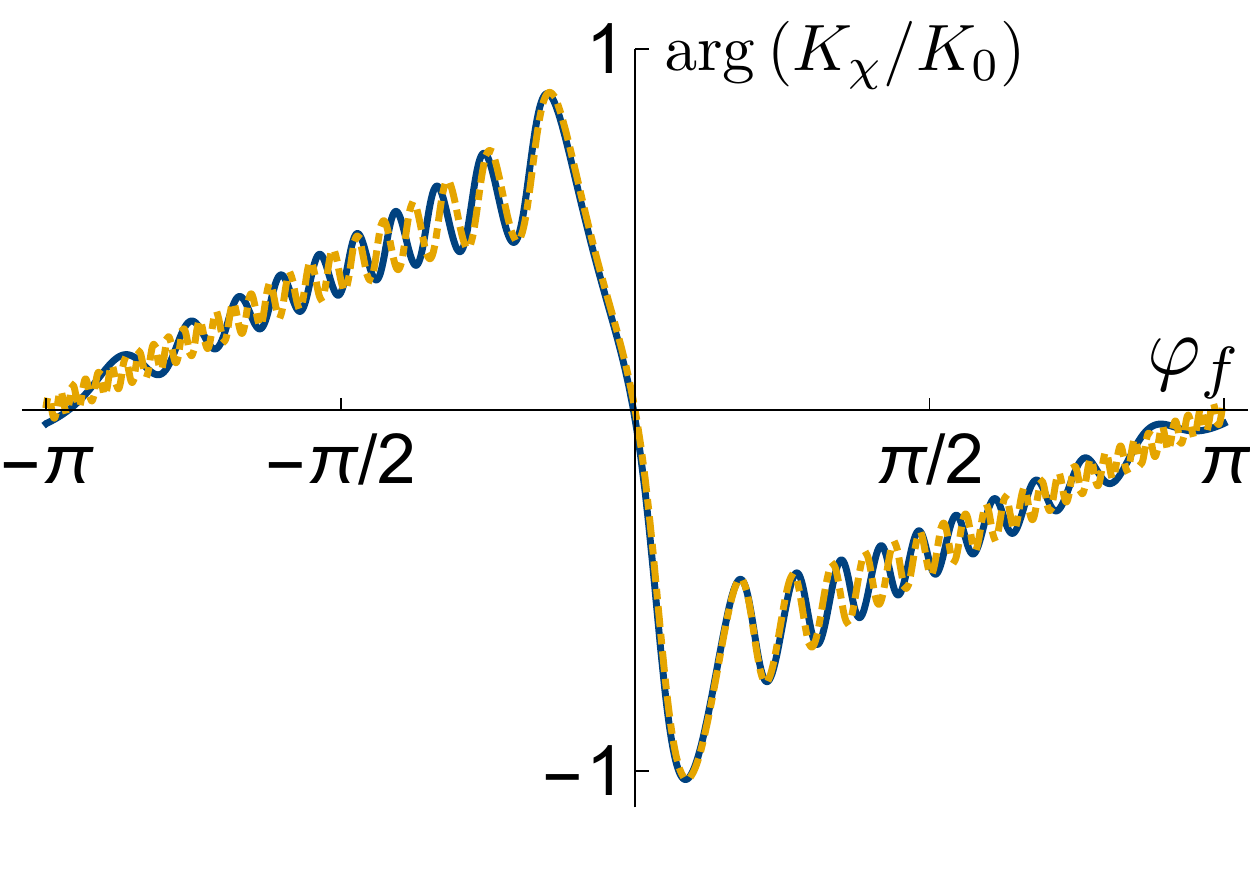}
  }
  \caption{Comparison of the normalized probability amplitude $|K_\chi|^2/|K_0|^2$ of the exact propagator~(\ref{eq:PropagatorIntegral}) (blue, full line) 
to that of
  the split-wave approximation~(\ref{eq:Kforward}) (yellow, dot-dashed), centered around the forward direction $\varphi_f=0$, for 
  $\chi/\hbar = 0.25$. 
 The agreement improves as  
  $z$ increases (a-c).  
 Analogous observations hold for the 
  normalized phase $\arg(K_\chi/K_0)$ (d). 
 }
  \label{fig:AharonovBohmInterference}
\end{figure}

\begin{figure}
\centering
  \subfloat[$z = 3\pi$\label{fig:backward3pi}]{%
    \includegraphics[width=0.46\textwidth]{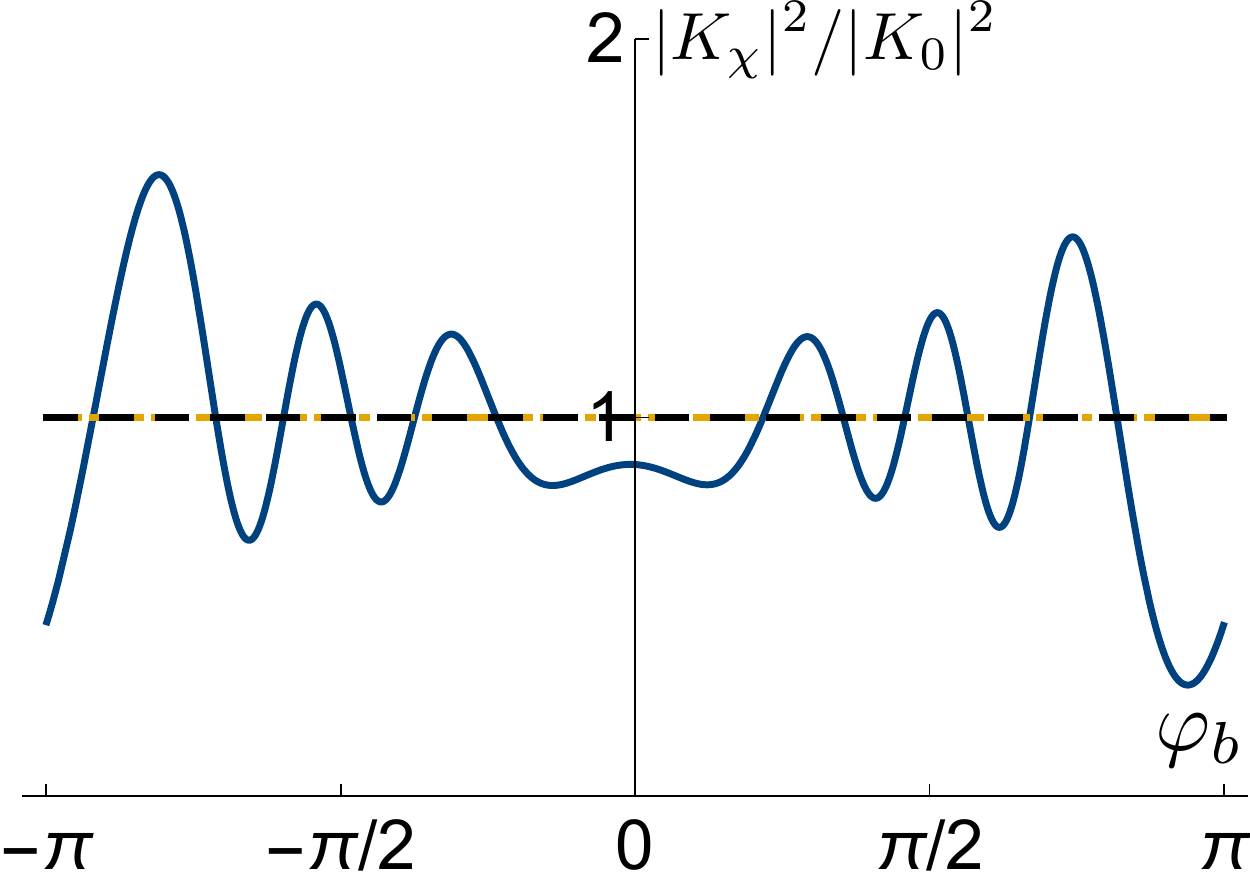}
  }
  \hfill
  \subfloat[$z =  3\pi \cdot 10^2$\label{fig:backward300pi}]{%
    \includegraphics[width=0.46\textwidth]{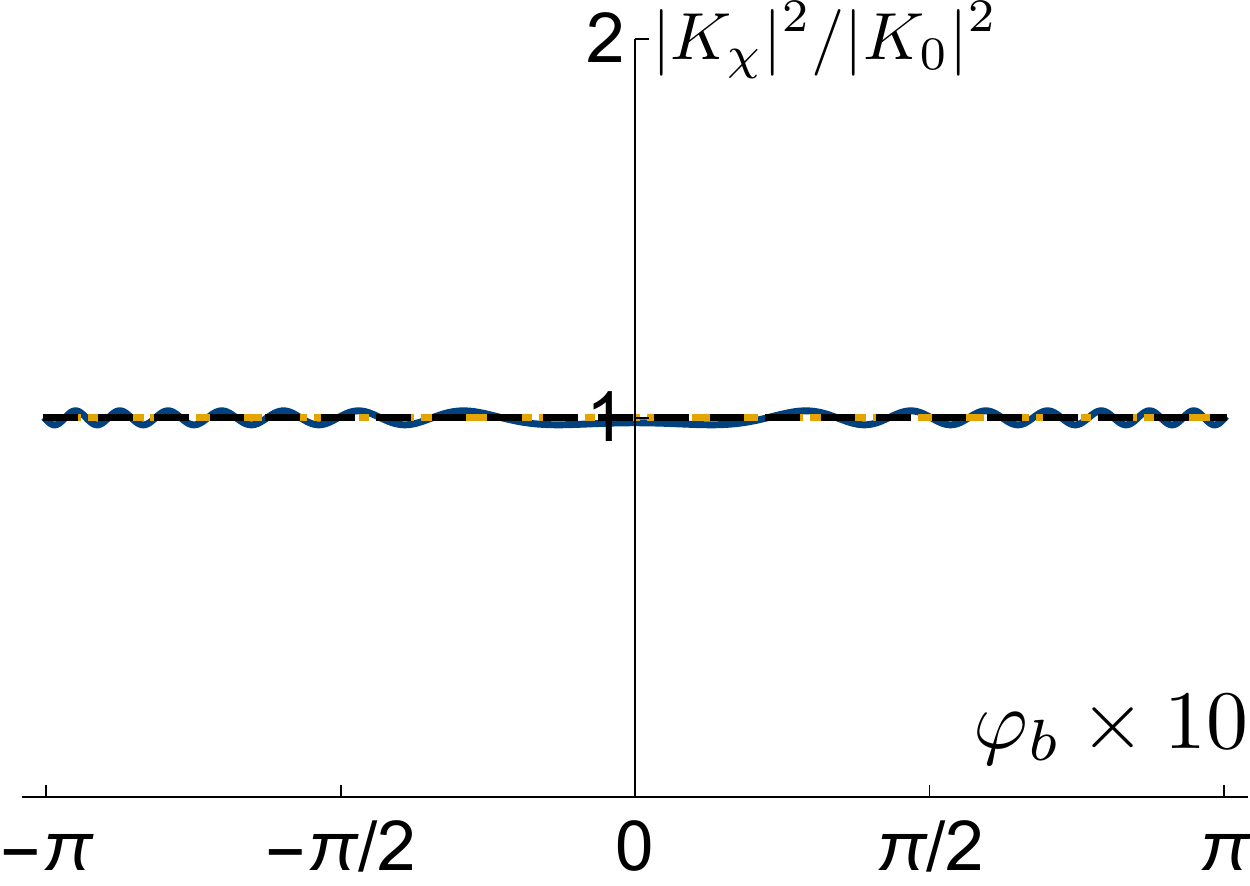}
  }
  
  \subfloat[$z = 3\pi \cdot 10^4$\label{fig:backward30000pi}]{%
    \includegraphics[width=0.46\textwidth]{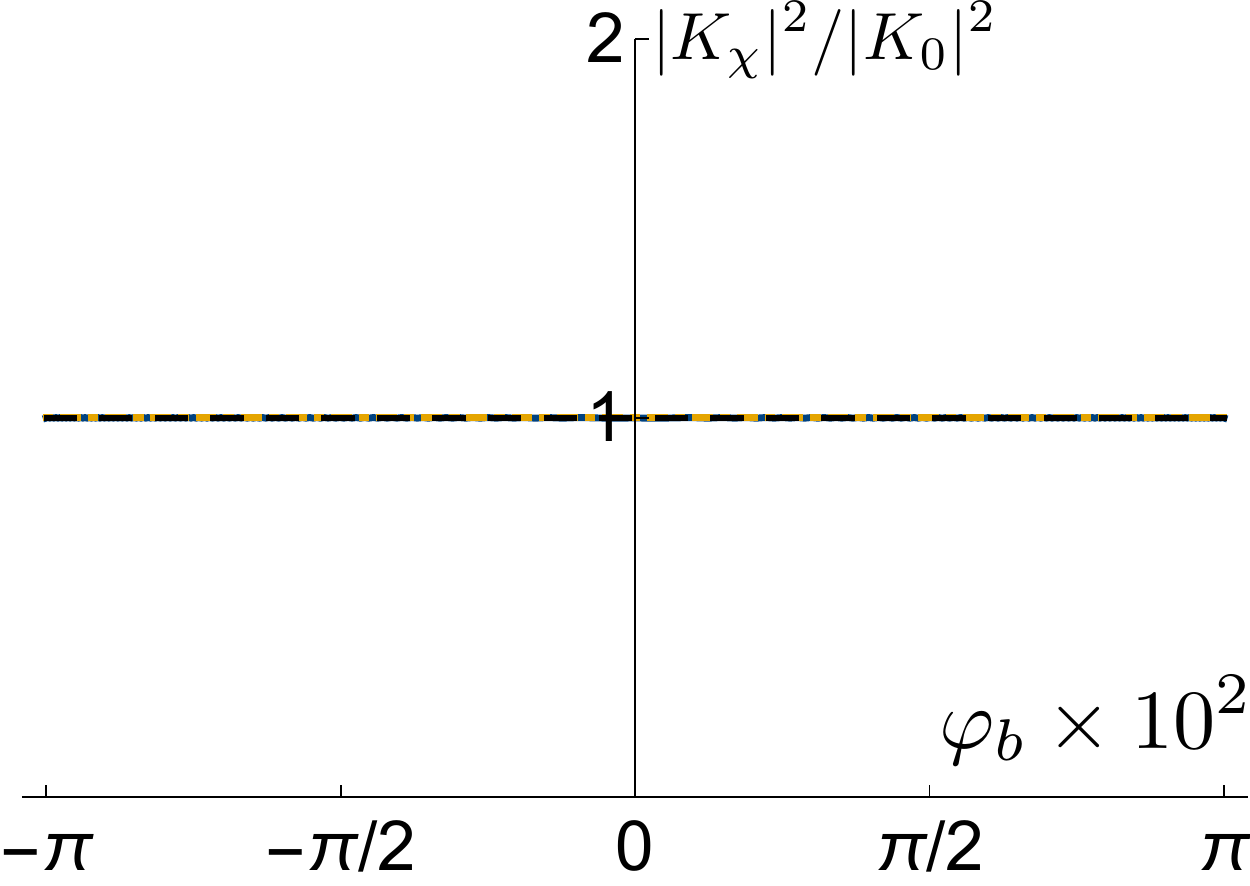}
  }
  \hfill
  \subfloat[$z = 10\pi $\label{fig:backward10pi}]{%
    \includegraphics[width=0.46\textwidth]{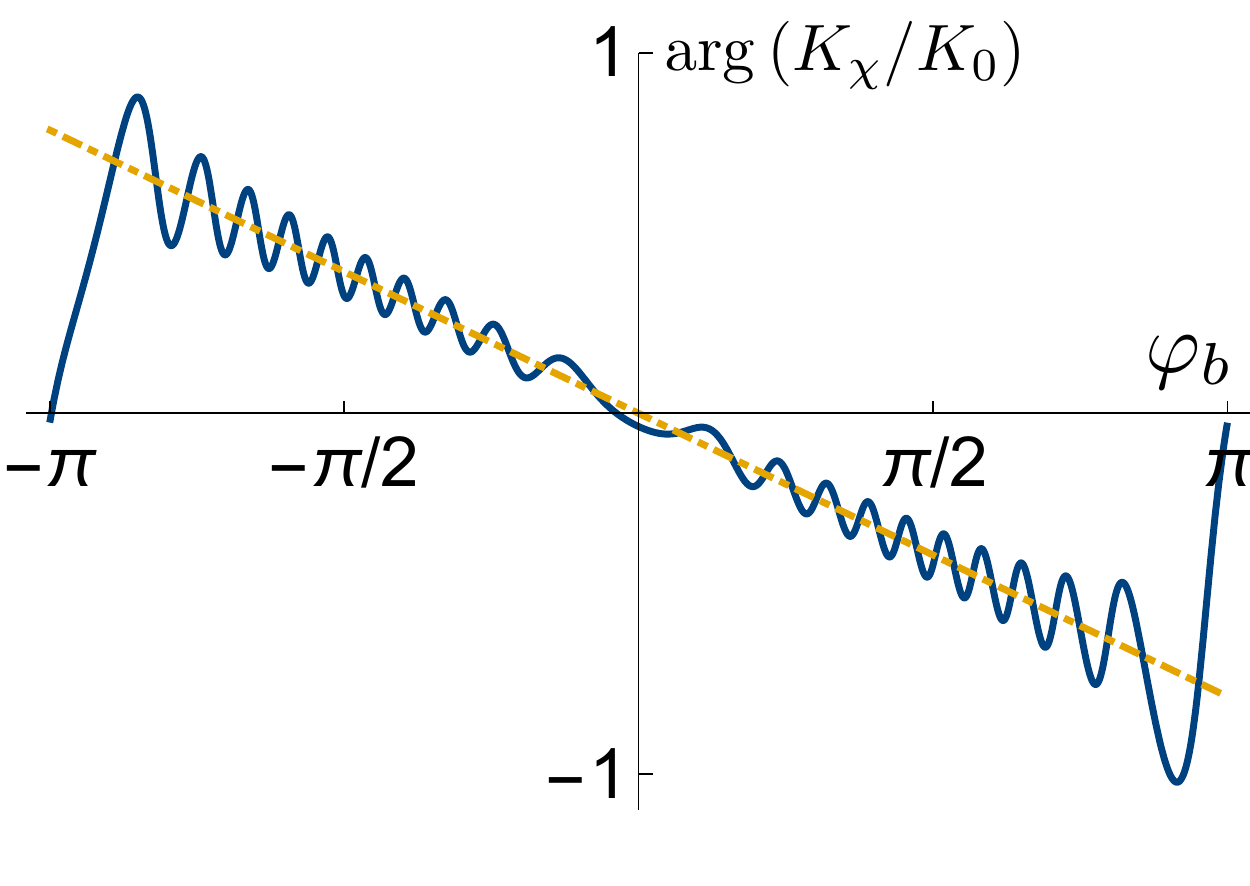}
  }
  \caption{Comparison of the normalized probability amplitude $|K_\chi|^2/|K_0|^2$ of the exact propagator~(\ref{eq:PropagatorIntegral}) (blue, full line) 
to that of
  the semiclassical (backward) approximation~(\ref{eq:absemprop}/\ref{eq:Kbackward}) (yellow, dot-dashed), centered around the backward direction $\varphi_b=0$, for 
  $\chi/\hbar = 0.25$. Comparable accuracy to the forward approximation (\ref{eq:Kforward}) is achieved only for larger $z$-values (c). 
 }
  \label{fig:AharonovBohmInterferenceBackwards}
\end{figure}

In figures~\ref{fig:AharonovBohmInterference} and~\ref{fig:AharonovBohmInterferenceBackwards}, we compare our forward~(\ref{eq:Kforward}) and backward~(\ref{eq:Kbackward}) approximations, respectively, to the integral representation~(\ref{eq:PropagatorIntegral}) of the exact propagator~(\ref{eq:prab}).
Figure~\ref{fig:AharonovBohmInterference} shows a comparison of the modulus squared of the full propagator~(\ref{eq:PropagatorIntegral}) to the semiclassical forward approximation~(\ref{eq:Kforward}), both divided by the modulus squared of the free-particle propagator, $|K_\chi|^2/|K_0|^2$. The expression yields the enhancement of the probability density to find a particle prepared at $\vec{x}'$ at position $\vec{x}''$ with respect to that of the free evolution, and depends only on three parameters: $z$, as defined in (\ref{eq:zphi}), increases with the distance of the preparation point $\vec{x}'$ and of the detection point $\vec{x}''$ from the solenoid, and decreases with increasing propagation time. $\varphi_f$ measures the angular distance from the forward direction. $\chi$ quantifies the flux.

Semiclassical and exact expression show good agreement around the forward direction $\varphi_f=0$ already for rather small values of $z$, and the accuracy of (\ref{eq:Kforward}) increases with $z$, see figure~\ref{fig:zh1}--\ref{fig:zh50}. The same applies to the normalized phase $\arg(K_\chi/K_0)$, see figure~\ref{fig:ph50}. Around the backward direction $\varphi_b=0$, comparable agreement between the exact propagator~(\ref{eq:PropagatorIntegral}) and the semiclassical approximation~(\ref{eq:absemprop},\ref{eq:Kbackward}) is achieved only for larger $z$ values (of the order $10^4 \pi$), whereat $|K_\chi|^2/|K_0|^2$ converges to unity, see figure~\ref{fig:AharonovBohmInterferenceBackwards}.


  \begin{figure}
  \centering
    \includegraphics[width=0.92\textwidth]{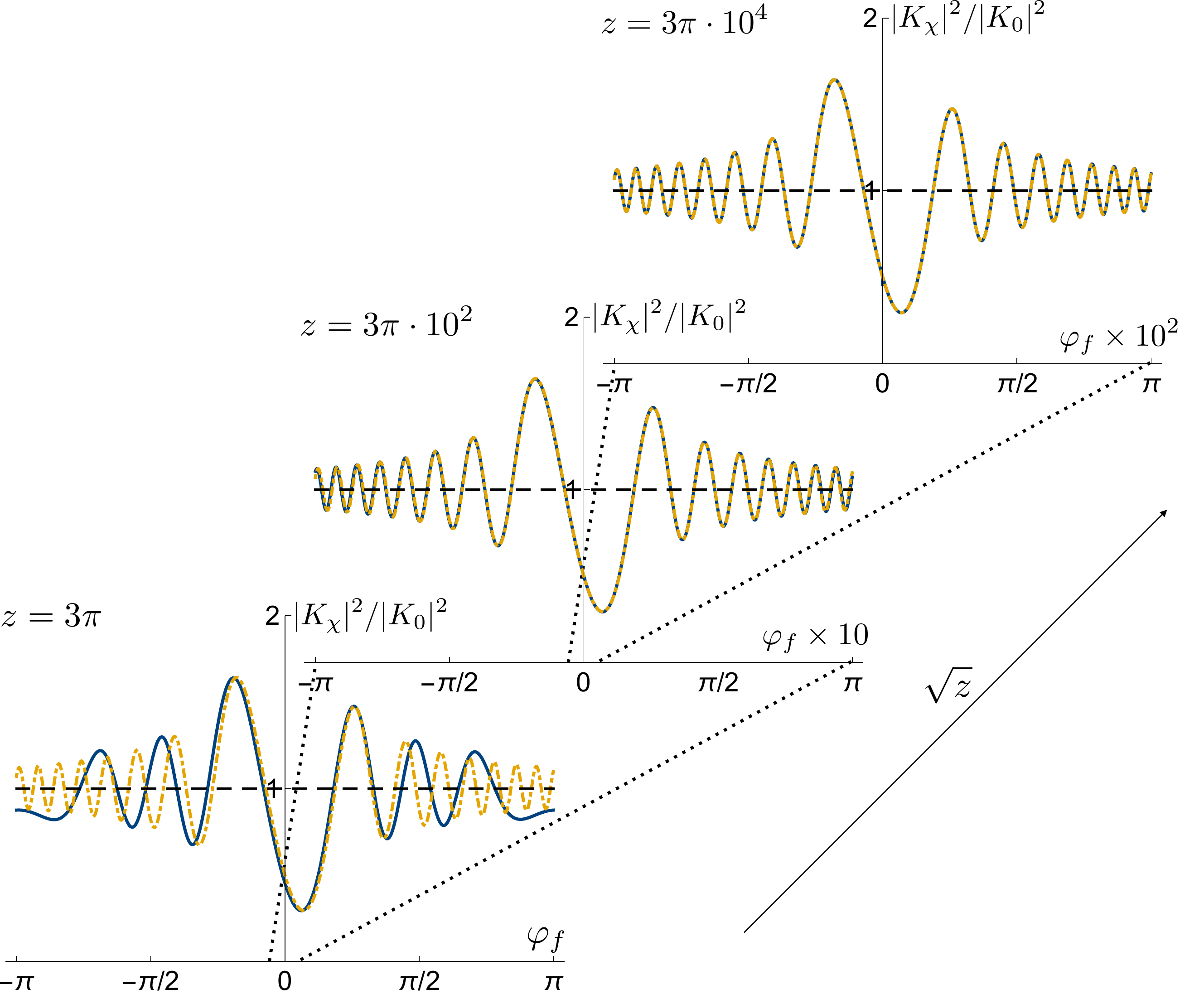}
  \caption{Whenever (\ref{eq:Kforward}) (yellow, dot-dashed) constitutes a suitable approximation to (\ref{eq:PropagatorIntegral}) (blue, full line), $|K_\chi|^2/|K_0|^2$ scales universally in the angular part $R_{\varphi_f}/\hbar = z \varphi_f^2 / 2$ of Hamilton's principal function in units of $\hbar$. 
  Increasing the square root of $z$ by an arbitrary amount contracts 
  the angular range of the interference pattern 
  by the corresponding factor. Ultimately, the width of the interference pattern 
   vanishes as $\sqrt{z}\rightarrow\infty$, giving rise to the discontinuity of (\ref{eq:absemprop}) in forward direction ($\chi/\hbar = 0.25$).
  }
  \label{fig:AharonovBohmScale}
\end{figure}

In the closing of this section we point out yet another property of the half-wave approximation~(\ref{eq:Kforward}), which facilitates the visualization of the Aharonov-Bohm propagator~(\ref{eq:prab}) in the asymptotic limit: The normalized probability density depends only on a product of the parameters $z$ and $\varphi_f$, which can be inferred neither from 
(\ref{eq:prab}) nor from (\ref{eq:PropagatorIntegral}). Thus, whenever (\ref{eq:Kforward}) is in good agreement with these expressions, we expect that the associated interference pattern scales universally in the parameter $z \varphi_f^2 / 2$ which, close to the forward direction, constitutes the angular contribution to Hamilton's principal function~(\ref{eq:HPF1}) of a free particle, 
\begin{align}
  R_{\varphi_f} = \frac{m}{2}  r'' r' \frac{\varphi_f^2}{t''-t'},
\end{align} 
in units of $\hbar$ (see also~(\ref{eq:Kforwardplus},\ref{eq:Kforwardminus})). Thus, if $z$ is multiplied by the square of an arbitrary factor, we expect the same interference pattern in an appropriately rescaled angular range. This is demonstrated in figure~\ref{fig:AharonovBohmScale} where we compare the prediction of (\ref{eq:PropagatorIntegral}) to the scale-invariant result  (\ref{eq:Kforward}), for several values of $z$.

The above implies that, as the width of the interference pattern decreases with the square root of $z$, and vanishes as $\sqrt{z}\rightarrow\infty$
(or, equivalenty,
$\sqrt\hbar\rightarrow 0$)\cite{BERRYII,OLARIUPOPESCU,SIEBER}, it does so in an invariant shape.
If $\varphi_f$ is positive, 
the bounds of the integrals in (\ref{eq:Kforward}) go to $-\infty$ as $\sqrt\hbar\rightarrow 0$,
such that the first integral yields unity while 
the second vanishes.
In turn, only the second integral survives for negative 
$\varphi_f$.
Therefore, in both cases only that contribution to the propagator remains which encircles the solenoid in the same rotational sense as the 
classical trajectory associated with semiclassical approximation~(\ref{eq:absemprop}). The latter is thereby recovered 
also in forward direction, exactly as 
upon evaluation of the integrals in (\ref{eq:Kforward0}) by the method of stationary phase, which thus corresponds to neglecting $\hbar$ against all finite values of $R_{\varphi_f}$. Finally, for $\varphi_f = 0$, (\ref{eq:KOP}) and~(\ref{eq:Kforward}) can be evaluated directly and 
yield, independently of the value of $z$,
\begin{align}
\label{eq:Kforward2}
  &K_\chi \sim \frac{m}{2\pi i  \hbar t} \exp \Bigg( \frac{i}{\hbar}  \frac{m}{2}\frac{\left(r'' + r'\right)^2 }{t''-t'}   \Bigg) \cos \left( \frac{\chi \pi}{\hbar} \right), 
\end{align}
as first observed in \cite{BERRYII}. 

Our above considerations demonstrate how phase jumps of the semiclassical approximation (\ref{eq:SemicalssicalPropagator}) occur entirely in the shadow region defined by the classical trajectories after direct passage through the magnetic string.
This is also the case, e.g., for the object wave in the geometrical shadow of the solenoid in the electron-holographic experiments of \cite{TONOMURA,TONOMURAII}.  

\section{\label{sec:Whirlingwave}Relation to the whirling-wave representation}

In this section we discuss the relation of our asymptotic expressions~(\ref{eq:Kbackward}) and~(\ref{eq:Kforward}) to the whirling-wave representation first introduced in reference~\cite{BERRYI},
\begin{align}
\label{eq:WhirlingWaves}
 K_\chi = \frac{m}{2\pi i  \hbar \left( t''-t' \right)} &\exp{ \left( \frac{i}{\hbar} \frac{m}{2}\frac{r''^2 +r'^2}{t''-t'} \right) } \nonumber \\
  &   \times \sum_{n=-\infty}^\infty \int_{-\infty}^{\infty} d\lambda \,  I_{\left| \lambda \right|}\left( - i z \right) e^{i {\left(\lambda - \chi/\hbar \right) \left( \varphi''-\varphi' + 2\pi n \right)}},
\end{align} 
which results from the exact Aharonov-Bohm propagator~(\ref{eq:prab}) upon invoking the Poisson summation formula~\cite{LIGHTHILL}. 
The $n$-th whirling wave
\begin{align}
\label{eq:OneWhirl}
  T_n \left(z,\varphi \right) = \int_{-\infty}^{\infty} d\lambda \,  I_{\left| \lambda \right|}\left( - i z \right) e^{i {\lambda \left( \varphi + 2\pi n \right)}},
\end{align}
 which corresponds to the $n$-th contribution to the sum in~(\ref{eq:WhirlingWaves}), thereby acquires Dirac's magnetic phase $\exp (- i\chi \left( \varphi''-\varphi' + 2\pi n \right)/\hbar)$. This property ensures the single-valuedness of~(\ref{eq:WhirlingWaves}) for arbitrary angular differences $\varphi''-\varphi'$~\cite{BERRYI}. For our comparison, we state here various aspects of the whirls~(\ref{eq:OneWhirl}) which, including their asymptotic behavior, have been extensively covered in reference~\cite{BERRYIII}.   
 
 \begin{figure}
\centering
  \subfloat[$z=\pi/4$\label{fig:WWNoLimit}]{%
    \includegraphics[width=0.55\textwidth]{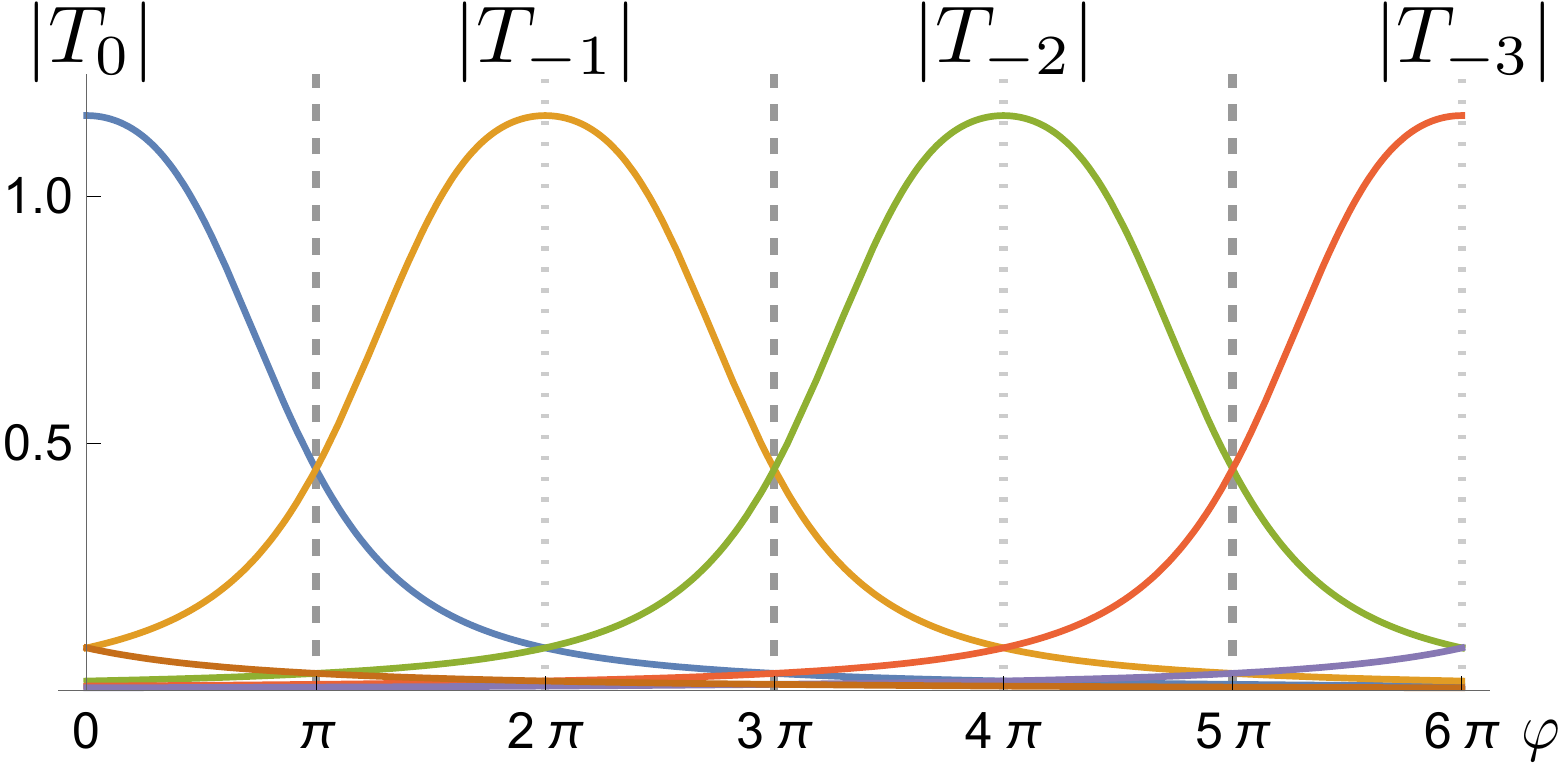}
  }
  
  \subfloat[$z=8\pi$\label{fig:WWAsymptoticLimit}]{%
    \includegraphics[width=0.55\textwidth]{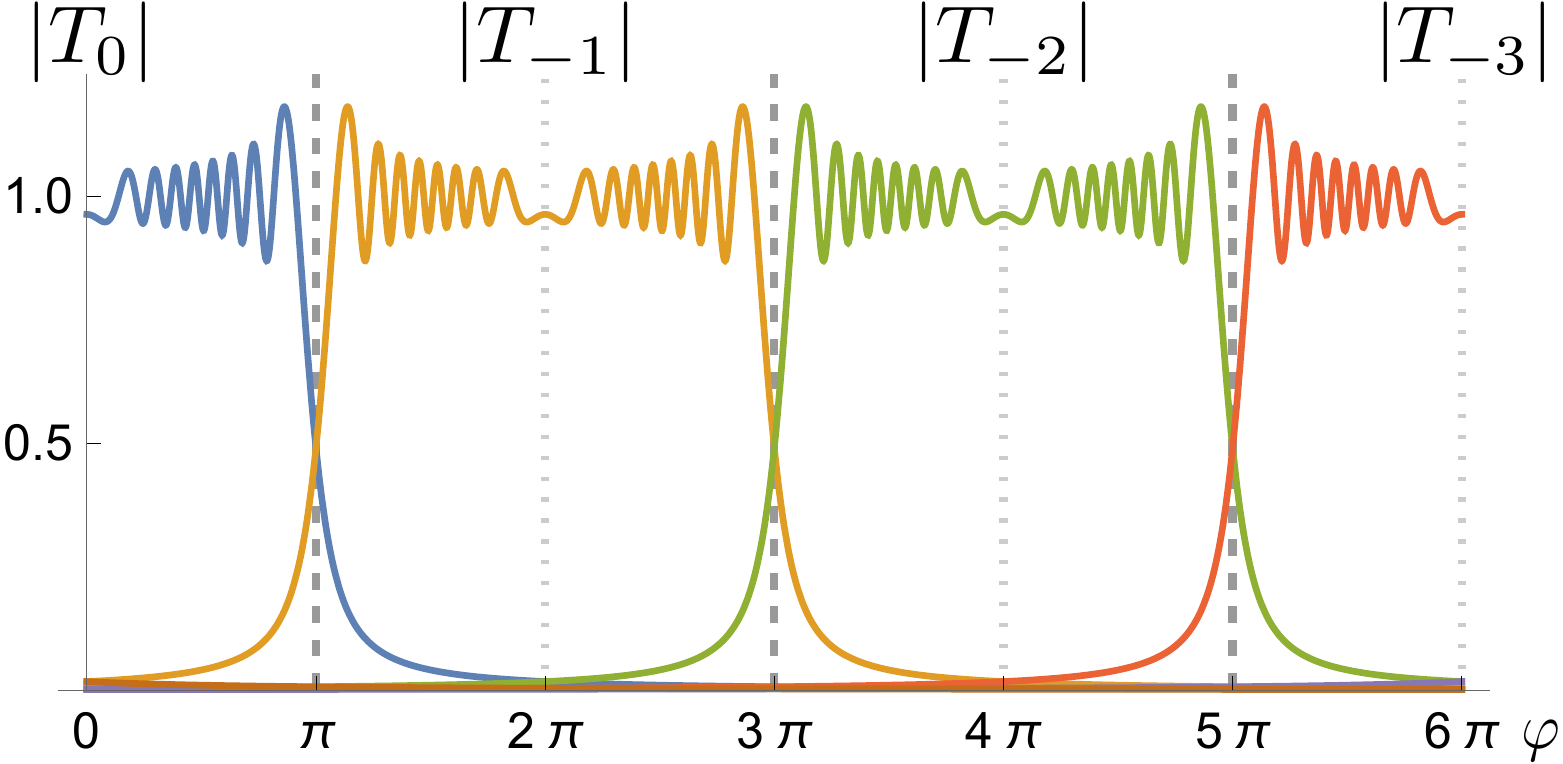}
  }
  
  \subfloat[$z=10^{10}\pi$\label{fig:WWSemiclassicalLimit}]{%
    \includegraphics[width=0.55\textwidth]{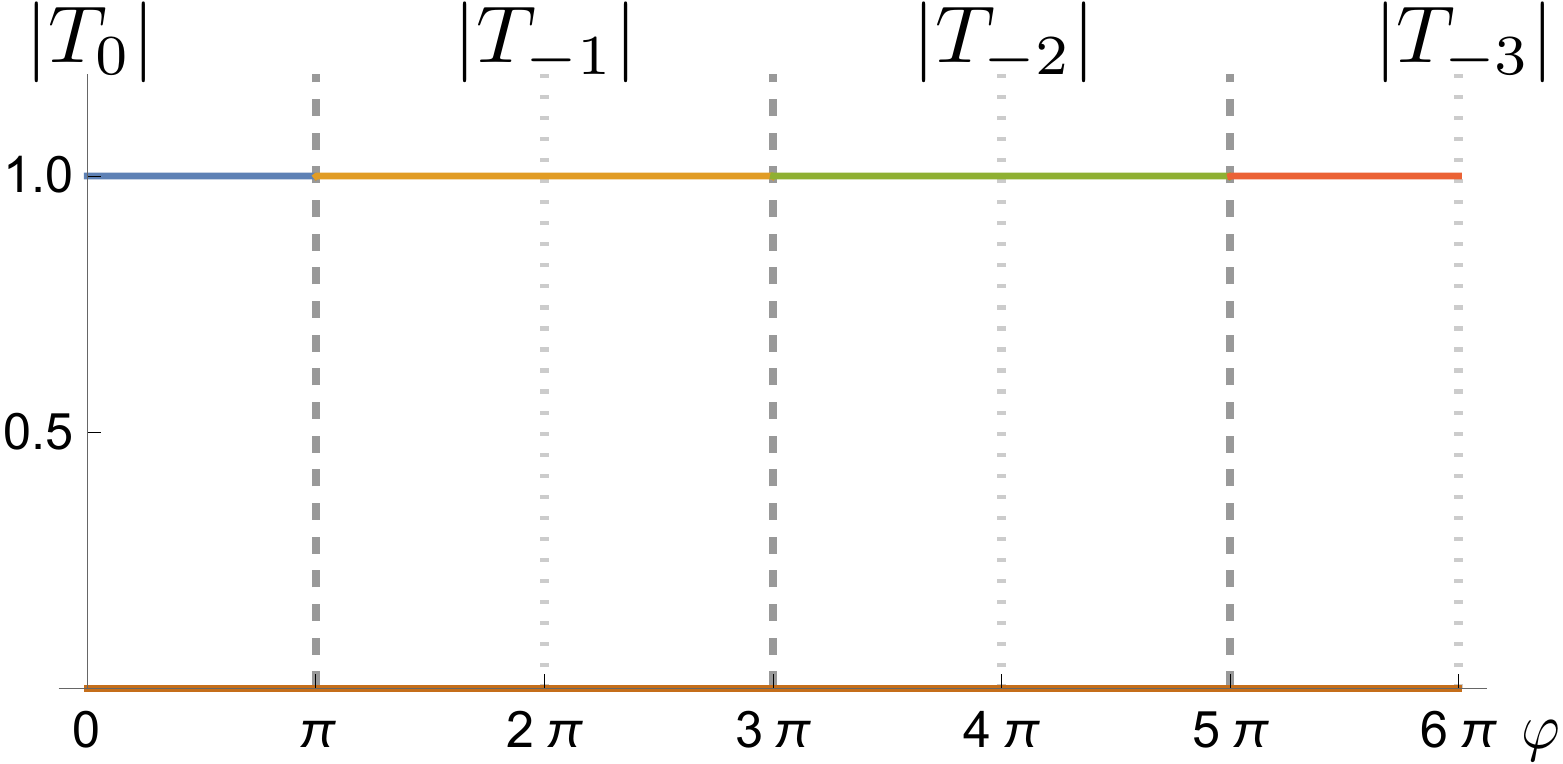}
  } 
    \caption{Amended for Dirac's magnetic phase, the whirling waves~(\ref{eq:OneWhirl}) in~(\ref{eq:WhirlingWaves}) give rise to an exact representation of the Aharonov-Bohm propagator~(\ref{eq:prab})~\cite{BERRYI}. (a,b) Each of the whirls is centered on a specific instance of the backward direction (gray dotted), and the tails of the whirls retract as $z$ increases~\cite{BERRYIII}. Our semiclassical limit~(\ref{eq:sumtoint}) shows that, asymptotically, only one whirl contributes in the backward direction, see~(\ref{eq:Kbackward}), and two whirls in the forward direction (gray dashed), see~(\ref{eq:Kforward}). This means that in this limit each whirl cannot extend beyond the backward direction of the respective adjacent whirls. (c) In the limit $z \to \infty$ the semiclassical propagator~(\ref{eq:absemprop}) is recovered, with the value~(\ref{eq:Kforward2}) in the forward direction (not shown).}
  \label{fig:WhirlingWaves}
\end{figure}  
 
  In figure~\ref{fig:WhirlingWaves} we plot several of the whirling waves, for different values of $z$. Each of the whirls~(\ref{eq:OneWhirl}) is centered on a particular instance of the backward direction at even integer multiples of $\pi$ (dotted vertical lines in figure~\ref{fig:WhirlingWaves}), and adjacent whirling waves give the largest contribution in the forward direction, at odd integer multiples of $\pi$ (dashed vertical lines in figure~\ref{fig:WhirlingWaves}). Thus it is common practice to suggest neglecting all but two~\cite{BERRYI,BERNIDOINOMATAII,BERNIDOINOMATAIII,MORANDIMENOSSI,BERRYIII} or three~\cite{BERRYIII} of the whirling waves, which constitutes a sound approximation of the exact Aharonov-Bohm propagator~(\ref{eq:prab}), in a specific angular range. Such approximations, however, feature a small discontinuity that cannot be avoided if any number of the whirling waves~(\ref{eq:OneWhirl}) are excluded from the sum in~(\ref{eq:WhirlingWaves})~\cite{BERRYIII}. Figures~\ref{fig:WWNoLimit} and~\ref{fig:WWAsymptoticLimit} show that for increasing $z$ the tails of the whirling waves retract, that is the overall contribution of each whirling wave increases in a $2\pi$ interval centered on its maximum and decreases outside of this interval~\cite{BERRYIII}. In the limit $z \to \infty$, the modulus squared of the whirling waves eventually converges to unity inside this interval, returning the value $1/2$ at the edges in the forward direction, and vanishing outside, see figure~\ref{fig:WWSemiclassicalLimit}. This corresponds to the semiclassical approximation~(\ref{eq:absemprop}) amended for the value~(\ref{eq:Kforward2}) in the forward direction $\varphi_f = 0$.

In fact, our semiclassical analysis in section~\ref{sec:III} above shows that in the backward direction the contribution of only one of the whirling waves remains, cf.~(\ref{eq:Kbackward}), and in the forward direction the same is true for only two of the whirling waves, cf.~(\ref{eq:Kforward}), in our semiclassical asymptotic limit~\footnote{This conclusion can also be drawn from the analysis in reference~\cite{BERRYVIII} which, in addition to the asymptotic limit, explicitly presupposes a paraxial incoming wave.}. Each whirling wave thus does not contribute beyond the backward direction of the respective adjacent whirls in this limit.
We therefore surmise that if an asymptotic approximation of~(\ref{eq:prab}) exists, smooth in the entire angular range and containing the contribution of only two of the whirls~(\ref{eq:OneWhirl}), asymptotically, one of these whirls must smoothly vanish as the backward direction is approached, while the other whirl must smoothly converge to coincide with the semiclassical approximation in the backward direction~(\ref{eq:Kbackward}).

For half-integer flux $\chi/\hbar$, even and odd contributions to the sum in~(\ref{eq:WhirlingWaves}) contribute with the same magnetic phase. Thus we find  
\begin{align}
\label{eq:WhirlingWavesHalfFlux}
 & K_\chi = \frac{m}{2\pi i  \hbar \left( t''-t' \right)} \exp{ \left( \frac{i}{\hbar} \frac{m}{2}\frac{r''^2 +r'^2}{t''-t'} \right) } \nonumber \\
  & \times \Bigg( \exp\left(-\frac{i}{\hbar} \chi \left(\varphi''- \varphi' \right)  \right) \sum_{n\, \mathrm{even}} \int_{-\infty}^{\infty} d\lambda \,  I_{\left| \lambda \right|}\left( - i z \right) e^{i {\lambda \left( \varphi''-\varphi' + 2\pi n \right)}} \nonumber \\
  & +  \exp\left(-\frac{i}{\hbar} \chi \left(\varphi''- \varphi' \mp 2\pi \right)  \right) \sum_{n\, \mathrm{odd}} \int_{-\infty}^{\infty} d\lambda \,  I_{\left| \lambda \right|}\left( - i z \right) e^{i {\lambda \left( \varphi''-\varphi' + 2\pi n \right)}} \Bigg),
\end{align} 
and indeed the first line of~(\ref{eq:WhirlingWavesHalfFlux}), together with the sum in the second line, yield the upper half-wave $K_0^+$ defined in~(\ref{eq:KzeroPlus}), while the first and third line, accordingly, yield the lower half-wave $K_0^-$~\footnote{For $K_0^+$, this can be seen by substituting $\lambda = \lambda'/2$ in~(\ref{eq:WhirlingWavesHalfFlux}), which after a subsequent reversal of the Poisson summation establishes equality to~(11.2.38) in~\cite{MORSEFESHBACH}. For $K_0^-$ this follows after the same steps which here entails a shift of $\varphi''-\varphi'$ by $2\pi$. }. Thereby, the connection to~(\ref{eq:semiflux}) is established, which shows why the exact  half-integer flux propagator already coincides with its semiclassically asymptotic form in the forward direction, compare~(\ref{eq:Kforward}), for arbitrary $z$-values.

\section{\label{sec:VI}Conclusions}

In order to address discontinuities which arise when the semiclassical approximation of Van Vleck/Gutzwiller is applied to approximately describe magnetic strings, we have introduced a novel semiclassical limit for the Aharonov-Bohm propagator. In this limit, the discrete sum over the quantum mechanical canonical angular momenta turns into integrals over a continuous variable, and the stationary points of those integrals coincide with the classical value of the canonical angular momentum. While our limit directly leads to the semiclassical approximation of Van Vleck/Gutzwiller in the backward direction, it generates in the forward direction an asymptotic split-wave expression of two contributions passing the string on either side. Thereby, the positive/negative kinetic angular momenta acquire Dirac's magnetic phase as if passing the string not more than once in a counterclockwise/clockwise rotational sense. We pointed out that the interference pattern generated by our split-wave approximation scales universally in the angular part of Hamilton's principal function in units of Planck's constant. Divergence of this ratio gives rise to the above-mentioned discontinuities, which corresponds to an evaluation of the canonical angular momentum integrals by means of the stationary phase approximation, and eventually leads to the expression of Van Vleck/Gutzwiller also in the forward direction. Our approach hints at features of a -- hypothetical -- globally smooth asymptotic approximation, whence we surmise that the form of such a function coincides in the respective regions with our asymptotic expressions for the forward and backward directions. We have shown that of a representation of the exact propagator in terms of whirling waves remain only one contribution in the backward direction and two contributions in the forward direction, in our limit.
In the case of half integer flux in which the propagator can be written in a closed form, every other whirl contributes with the same magnetic phase factor, directly giving rise to one of the half-waves which are for an arbitrary value of the flux encountered in the forward direction only in the asymptotic limit.

\section*{Acknowledgments}

The authors would like to thank Michael V.~Berry for a helpful discussion and comments.

\newpage
\bibliography{PaperNJP}

\end{document}